\documentclass[preprint,12pt]{elsarticle}

\usepackage{amsmath,amsfonts,amsthm,amssymb,paralist,subfigure,graphicx,amsbsy,float,epsfig,color}
\usepackage{cuted,mathtools,lipsum}

\usepackage{algorithm,algpseudocode}
\usepackage{hyperref}       % hyperlinks
\usepackage{stmaryrd,url}

\usepackage{amsthm}

\def\mb{\mathbf}

\def\mc{\mathcal}

\DeclareMathOperator*{\argmin}{argmin}
\DeclareMathOperator*{\argmax}{argmax}

\journal{Annual Reviews in Control}

\begin{document}

\begin{frontmatter}



\title{Survey of Distributed Algorithms for Resource Allocation over Multi-Agent Systems
%: A Survey of Distributed Optimization subject to Coupling Equality-Constraint
}


\author[Sem]{Mohammadreza Doostmohammadian}
\affiliation[Sem]{Faculty of Mechanical Engineering, Semnan University, Semnan, Iran. Email: doost@semnan.ac.ir.}
\author[AA]{ Alireza Aghasi} %\ead{aaghasi@gsu.edu},
\affiliation[AA]{Department of Electrical Engineering and Computer Science, Oregon
State University, USA. Email: alireza.aghasi@oregonstate.edu.}
\author[MP]{Mohammad Pirani}%\ead{rabiee@sharif.edu},
\affiliation[MP]{Department of Mechanical Engineering,
	University of Ottawa
	Ottawa, ON, Canada. Email: mpirani@uottawa.ca.}
\author[EN]{Ehsan Nekouei}%\ead{rabiee@sharif.edu},
\affiliation[EN]{City University of Hong Kong, Hong-Kong. Email: enekouei@cityu.edu.hk.}
\author[HZ]{ Houman Zarrabi}
\affiliation[HZ]{Iran Telecom Research Center (ITRC), Tehran, Iran. Email: h.zarrabi@itrc.ac.ir.}
\author[RK]{ Reza Keypour}
\affiliation[RK]{Faculty of Electrical Engineering, Semnan University, Semnan, Iran. Email: rkeypour@semnan.ac.ir.}
%%
% \author[َAR]{Apostolos I. Rikos}
% \affiliation[AR]{Division of Decision and Control Systems, KTH Royal Institute of Technology, Sweden, rikos@kth.se}
%%
\author[AR]{Apostolos I. Rikos}
\affiliation[AR]{Department of Electrical and Computer Engineering, Division of Systems Engineering, Boston University, Boston, MA, USA. Email: arikos@bu.edu.}
\author[KJ]{Karl H. Johansson}
\affiliation[KJ]{Division of Decision and Control Systems, KTH Royal Institute of Technology, Sweden. Email: kallej@kth.se}

\begin{abstract}
	Resource allocation and scheduling in multi-agent systems present challenges due to complex interactions and decentralization. This survey paper provides a comprehensive analysis of distributed algorithms for addressing the distributed resource allocation (DRA) problem over multi-agent systems. It covers a significant area of research at the intersection of optimization, multi-agent systems, and distributed consensus-based computing. The paper begins by presenting a mathematical formulation of the DRA problem, establishing a solid foundation for further exploration. Real-world applications of DRA in various domains are examined to underscore the importance of efficient resource allocation, and relevant distributed optimization formulations are presented.
	The survey then delves into existing solutions for DRA, encompassing linear, nonlinear, primal-based, and dual-formulation-based approaches.
	Furthermore, this paper evaluates the features and properties of DRA algorithms, addressing key aspects such as feasibility, convergence rate, and network reliability.
	The analysis of mathematical foundations, diverse applications, existing solutions, and algorithmic properties contributes to a broader comprehension of the challenges and potential solutions for this domain.
	
\end{abstract}

\begin{graphicalabstract}
	\includegraphics{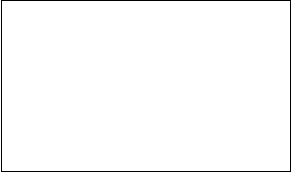}
\end{graphicalabstract}

\begin{highlights}
	\item Clearly defining the problem of resource allocation and scheduling in multi-agent systems and presenting a formal mathematical framework that captures its essence.
	\item Exploring various real-world applications where resource allocation and scheduling are crucial in cloud computing, transportation systems, and smart grids.
	\item Summarizing state-of-the-art solutions for distributed resource allocation, including linear, nonlinear, primal-based, and dual-formulation-based approaches.
	\item Analyzing different features of the existing algorithms in terms of feasibility, convergence, and network reliability.
\end{highlights}


\end{frontmatter}


\section{Introduction} \label{sec_intro}
In today's technological landscape, decentralized architectures are on the rise. Distributed algorithms are gaining increasing importance, thanks to their compatibility with various distributed platforms like the Internet of Things (IoT), cloud-based computing, and parallel processors. Many real-world systems, such as computer networks, peer-to-peer networks, cloud computing environments, and blockchain networks, inherently operate in a distributed fashion. Distributed algorithms serve as the cornerstone for designing efficient and scalable solutions in these systems. By studying and developing these algorithms, we can address practical challenges, enhance system performance, and improve functionality. Furthermore, distributed algorithms offer fault-tolerant solutions by distributing workload and data across multiple nodes/agents\footnote{In this paper, we use node and agent interchangeably.}, mitigating the single-point-of-failure issue. In the event of a node failure or malfunction, the system can properly redistribute tasks to the remaining nodes, ensuring continued operation. This fault-tolerance property ensures that the system remains operational and resilient in the face of failures, leading to increased reliability and availability \cite{dibaji2019systems}. The large-scale applications range from distributed estimation and monitoring of dynamical systems  \cite{jstsp,kar2013consensus,nuno-suff.ness}, distributed coordination among robots for cooperative tasks  \cite{jadbabaie2003coordination}, fault-detection in dynamical systems \cite{cps_bullo} to more recent optimization, machine learning and distributed data-mining solutions  \cite{khan2020optimization} and power flow optimization in smart grids \cite{molzahn2017survey,yang2013consensus,themis2011asynchronous,cherukuri2015distributed}.

Distributed resource allocation (DRA) algorithms are designed to efficiently distribute limited resources among multiple entities or agents in a decentralized manner. These algorithms aim to optimize the allocation process by considering factors such as fairness (cost-optimality), efficiency, and scalability. The motivation behind these algorithms can be understood through several key aspects:
\begin{itemize}
\item \textit{Scalability:} In many real-world scenarios, there are large-scale systems with numerous entities competing for limited resources. Centralized approaches may struggle to handle the computational and communication overhead to the central processor required to manage such systems. DRA algorithms address this challenge by allowing entities to make local decisions based on limited information, thus enabling scalable resource allocation in large networks \cite{pelikan2006scalable}.
\item \textit{Efficiency:} Efficient utilization of resources is crucial to maximizing the overall system performance. By distributing the decision-making process across multiple entities, distributed algorithms can exploit local knowledge and adapt to dynamic conditions, leading to more efficient resource allocation \cite{vinyals2011survey}. By using multiple nodes, distributed algorithms can make better use of available resources, as each node can leverage its \textit{local} resources more efficiently. Furthermore, distributed algorithms can facilitate decentralization and parallelizability, which is more efficient in scenarios where a central point of control may not be practical, cost-effective, or desirable, and tasks can be divided into smaller independent subtasks and processed in parallel. Examples include peer-to-peer networks and block-chain systems \cite{ramseyer2023speedex}.
\item \textit{Flexibility and Adaptability:} DRA algorithms are designed to be flexible and adaptable to changing environments without the intervention of a centralized entity. They can locally respond to fluctuations in resource availability or changes in agent requirements by dynamically reallocating resources \cite{skobelev2015multi}. This adaptability is particularly valuable in dynamic and unpredictable systems, where the allocation requirements may vary over time.
%\item \textbf{Fairness:} Fairness is a critical consideration in resource allocation scenarios, as it ensures that resources are allocated in an equitable and cost-optimal manner. Fairness is typically defined via the objective function. Distributed algorithms can incorporate fairness criteria into their decision-making process, aiming to distribute resources fairly among the participating entities. Fairness metrics can vary depending on the application domain, and the algorithms can be customized accordingly to meet specific fairness objectives.
\item \textit{Fault-tolerance and Robustness:} Distributed systems are often prone to failures or disruptions due to the distributed nature of the environment \cite{shreyas2020byzantine,pirani2018cooperative}. DRA algorithms can be designed to be fault-tolerant and robust, allowing them to continue functioning even in the presence of failures or network partitions. These algorithms may incorporate redundancy, consensus mechanisms, or recovery strategies to ensure reliable resource allocation.
\item \textit{Privacy and Security:} In certain scenarios, entities may have privacy concerns or security requirements that limit the sharing of sensitive information \cite{pequito2014design,rikos_quant,ruan2019secure,mo2016privacy,nekouei2019information}. DRA algorithms can accommodate such constraints by allowing agents to make local decisions based on their own private information or a limited set of shared information. The decentralized approach reduces the exposure of sensitive data and enhances the security and privacy of the resource allocation process \cite{dsvm}.
\end{itemize}

The above-mentioned motivational aspects behind DRA algorithms lie in their ability to address the challenges posed by large-scale, dynamic, and decentralized systems. By adopting local decision-making, adaptability, and feasibility considerations these algorithms achieve efficient and effective resource allocation in a decentralized manner. These considerations further motivate many use-cases of DRA algorithms including wireless communication networks \cite{heinzelman2000energy,akyildiz2002survey}, distributed and cloud computing \cite{vlahakis2021aimd}, traffic management \cite{baldacci2012recent,taillard1996vehicle,gao2021priority}, power distribution and energy management \cite{boqiang2009review}, task assignment in multi-agent systems \cite{MiadCons}, radio resource management in cellular networks \cite{zander1997radio}, distributed sensor networks \cite{lesser2003distributed}, supply chain management \cite{malairajan2013class}, etc.

This work provides a comprehensive overview of the field of DRA and scheduling over multi-agent systems, and contributes to the understanding of challenges, existing solutions, and properties of such algorithms. First and foremost, to lay a solid foundation for the subsequent analysis, a mathematical formulation of the DRA problem is presented. This formulation captures the essence of resource allocation challenges in multi-agent systems and enables a rigorous examination of potential solutions. By defining the problem in mathematical terms, along with extensive preliminaries on graph theory, consensus algorithms, and convex set analysis, our paper sets the stage for a systematic exploration of applicable distributed algorithms. Furthermore, it provides a comprehensive and up-to-date survey of the various distributed optimization techniques utilized to address the challenges posed by coupling equality constraints in multi-agent systems. Furthermore, this survey explores existing solutions and applications for the DRA problem, focusing on smart grid (automatic generation control and economic dispatch), CPU scheduling over networked data centres,  plug-in electric vehicle (PEV) charging schedules, and network utility maximization (NUM). Additionally, both Laplacian-gradient-based solutions to the primal formulation and dual-formulation-based solutions are presented. A comprehensive comparative analysis is conducted, highlighting the strengths, limitations, and applicability of each algorithm to different scenarios. This examination not only presents the state-of-the-art in DRA but also identifies the key algorithmic techniques and principles employed in these solutions. Furthermore, the paper evaluates the main features and properties of the DRA algorithms: feasibility, convergence, and network reliability. Feasibility analysis assesses the algorithms' ability to satisfy the resource-demand coupling-constraint in all iterations. Convergence analysis examines how quickly these algorithms converge to optimal or near-optimal solutions. Network reliability considerations address the robustness of algorithms in the face of communication failures, dynamic changes in the network, latency, and packet drops.

As compared to the existing literature focusing solely on distributed optimization \cite{nedic2018distributed,yang2019distributed,molzahn2017survey}, this work is more devoted to specifically highlighting applications and formulations of the DRA algorithms. While both types of algorithms operate in a decentralized manner, general distributed optimization algorithms focus on solving global optimization problems, and DRA algorithms concentrate on efficiently allocating shared resources among multiple nodes. The choice between the two depends on the specific requirements and objectives of the distributed system in question. DRA algorithms are primarily concerned with dividing and allocating resources efficiently among multiple nodes in a distributed system with the help of \textit{constrained} optimization algorithms. The resources could include bandwidth, computing power, storage, or any other shared resource. The goal is to ensure optimal distribution or utilization and preserve constraint-feasibility among the nodes. Therefore, applications of the DRA algorithms are often different from those of distributed optimization. In contrast to mainly machine learning applications of distributed optimization in the literature, the DRA mainly focuses on computer networks, cloud computing environments, wireless communication systems, smart grid and energy networks where efficient resource allocation is essential for maintaining system stability and performance.

By presenting a comprehensive analysis of distributed algorithms for resource allocation and scheduling over multi-agent systems, this survey paper contributes to a deeper understanding of the field. The exploration of mathematical foundations, real-world applications, existing solutions, and algorithmic properties serves as a valuable resource for researchers, practitioners, and system designers seeking to optimize resource allocation in decentralized settings.

\textit{Paper Outline:} Section~\ref{sec_prob} states the DRA problem in mathematical form and Section~\ref{sec_pre} provides the preliminaries and background.  Section~\ref{sec_sol} discusses different existing distributed solutions to the DRA problem and in Section~\ref{sec_feature} the main features and properties of these solutions are discussed. Section~\ref{sec_app} presents real-world applications. Finally, Section~\ref{sec_conclusion} concludes the paper with some future directions.

\textit{General Notation:}
The column vectors, scalars, and matrices are respectively represented by bold small letters,  small letters,  and capital letters. $\partial f_i$ and $\partial^2 f_i$ denote the first derivative $ \frac{df_i(x_i)}{dx_i}$ and second derivative $ \frac{d^2f_i(x_i)}{dx_i^2}$, respectively. $\nabla F(\cdot)$  denotes the gradient of $F(\cdot)$.
Vectors $\mb{1}_n$, $\mb{0}_n$ denote column vector of ones and zeros of size $n$.

\section{The DRA Problem} \label{sec_prob}
\subsection{Motivational Example}
Consider a network of renewable energy sources, such as wind turbines and solar panels, scattered across a vast landscape. These sources generate clean, green energy, but their output is inherently intermittent and dependent on weather conditions. To maximize the utilization of these resources and ensure a reliable energy supply, a robust and adaptive resource allocation system is crucial.
Traditional power grids that rely on centralized decision-making struggle to efficiently integrate renewable energy sources. Grid operators had to predict energy generation, allocate resources based on these predictions, and often relied on fossil fuel backup when renewable sources fell short, resulting in wasted energy and increased greenhouse gas emissions.
Distributed resource allocation, on the other hand, introduces a game-changing approach. Each renewable energy source, equipped with sensors and intelligent control systems, can autonomously adjust its output in response to real-time weather data and energy demand. When one area experiences a sudden increase in energy demand due to, for example, an unexpected heatwave, nearby solar panels can increase their energy production, while wind turbines in windy regions can scale back during low wind forces. This decentralized decision-making ensures that energy is generated where it's needed most and minimizes energy loss during transmission.
Furthermore, the surplus energy generated during optimal conditions can be properly redistributed to areas with higher demand, enhancing grid resilience and reducing the need for backup power sources.
%The entire system adapts dynamically, harnessing renewable resources efficiently while reducing the environmental impact and reliance on fossil fuels.
Other advantages include scalability and not relying on single point of failure. Scalability implies that the algorithm is able to keep its efficiency when the system size and workload grow. No single point of failure implies that in case of losing one (or more) node/agent the rest of the networked system keeps working and performs its intended tasks (by relying on node redundancy).

Potential applications of the DRA extend beyond energy grids to various domains, including the allocation of computing resources in large-scale data centres, transportation systems, and communication networks. Such motivating applications are discussed in detail in Section~\ref{sec_app}.

\subsection{Mathematical Formulation}
In DRA, the idea is to update the state of some agents $z_i$ over a network to minimize their cost function (or certain loss function), while the weighted sum of states is constant and equal to certain demand $b$.
DRA in its most general form is modelled as an
equality-constraint coupled optimization problem as follows,
\begin{equation} \label{eq_dra0}
\begin{aligned}
	\displaystyle
	\min_\mb{z}
	~~ & \widetilde{F}(\mb{z}) := \sum_{i=1}^{n} \widetilde{f}_i(z_i) \\
	\text{s.t.} ~~&  \mb{z}^\top\mb{a} = b, ~~ \underline{z}_i \leq z_i \leq \overline{z}_i
\end{aligned}
\end{equation}
with $z_i \in \mathbb{R}$ as the state variable at agent $i$,
column vector $\mb{z} = [z_1;\dots;z_n] \in \mathbb{R}^{n}$ as the collective vector state\footnote{
The most general form of
the problem extends to $\mb{z}_i \in \mathbb{R}^m$ with $m>1$ as in \cite{doostmohammadian20211st,lakshmanan2008decentralized}. For the sake of simplicity and clarity of the problem formulation, here we assume $m=1$.},
and constraint parameters $\mb{a}=[{a}_1;\dots;{a}_n] \in \mathbb{R}^n$ and $b \in \mathbb{R}$. The function $\widetilde{f}_i: \mathbb{R} \mapsto  \mathbb{R}$ denotes the local objective at agent $i$, and the overall cost is $\widetilde{F}: \mathbb{R}^{n} \mapsto  \mathbb{R}$.
The box constraints $\underline{z}_i \leq z_i \leq \overline{z}_i$ are the local (non-coupling) constraints and $\mb{z}^\top\mb{a} = b$  is the coupling constraint.

Simplifying the problem described in \eqref{eq_dra0}, we employ a change of variables by introducing $x_i := z_i a_i$. This transformation allows us to express the problem in a different standard form commonly found in the existing literature. The simplified formulation of \eqref{eq_dra0} is as follows:
\begin{equation} \label{eq_dra}
\begin{aligned}
	\displaystyle
	\min_\mb{x}
	~~ & F(\mb{x}) = \sum_{i=1}^{n} f_i(x_i) \\
	\text{s.t.} ~~&  \sum_{i=1}^{n} x_i = b, ~~ \underline{x}_i \leq x_i \leq \overline{x}_i
\end{aligned}
\end{equation}
with $\underline{x}_i:=a_i \underline{z}_i$, $\overline{x}_i := a_i  \overline{z}_i$ for $a_i>0$ and reversed otherwise. The form for the DRA problem presented in \eqref{eq_dra} is the main focus of our paper.
The formulation in \eqref{eq_dra} involves the minimization of a specific objective function, such as cost or loss, while ensuring a balance in resource demand through the equality constraint $\sum_{i=1}^{n} x_i = b$. In simpler terms, the optimal solution necessitates that the total resources, represented by $\sum_{i=1}^{n} x_i$, align precisely with the overall demand $b$.
The box constraints $\underline{x}_i \leq x_i \leq \overline{x}_i$ imply that the states are upper and lower bounded. This is typically addressed by the barrier functions or penalty terms in the objective function. In other words, one can approximate these local constraints by additive penalty terms to local objective functions. Some example penalty functions are given in the literature \cite{bertsekas1975necessary,nesterov1998introductory,doostmohammadian20211st,mikael2021cdc}. One well-known case of penalty function is,
\begin{equation} \label{eq_penalty}
f_i^{\sigma} =  c([z_i - \overline{z}_i]^+ + [\underline{z}_i - z_i ]^+),
\end{equation}
with $[u]^+=\max \{u, 0\}^\sigma,~\sigma \in \mathbb{N}$, and $c \in \mathbb{R}^+$.
One can show that for sufficiently large $c$ the solutions of the penalized problem can get arbitrarily close to the solution of the problem \eqref{eq_dra}.
The penalizing terms (or barrier functions) are generally non-quadratic, see examples in \cite{doostmohammadian20211st,mikael2021cdc}.

Among the properties of the DRA solution, feasibility is a main concern. Mathematically, we define the set $\mc{S}_b = \{\mb{x} \in \mathbb{R}^{n}|\mb{x}^\top\mb{1}_n = b\}$ for problem \eqref{eq_dra} (or  $\widetilde{\mc{S}}_b = \{\mb{z} \in \mathbb{R}^{n}|\mb{z}^\top\mb{a} = b\}$ for problem \eqref{eq_dra0}) as the feasible set and $\mb{x} \in \mc{S}_b$ (or $\mb{z} \in \widetilde{\mc{S}}_b$) as the feasible solution. A feasible solution guarantees that the resource-demand constraint (the coupling constraint) is satisfied and there is no gap between the assigned resources and the demand. If a solution is all-time feasible, it is ensured that at any stopping time of the optimization algorithm, the constraints hold. Note that, under the local box constraints, feasible initialization algorithms are needed and are given, for example, in \cite{cherukuri2015distributed}. The feasibility property is more discussed in Section~\ref{sec_feas}.

%For the mentioned box constraints after transformation $z_i a_i = x_i$ the constraint changes into $a_i \underline{z}_i \leq x_i \leq a_i  \overline{z}_i$ for $a_i>0$ and $a_i \underline{z}_i \geq x_i \geq a_i  \overline{z}_i$ for $a_i<0$, which are convex in both cases\footnote{In this paper, we only consider the local box constraints in the form $\underline{x}_i \leq x_i \leq \overline{x}_i$. In the presence of more complex box constraints $h_i(x_i) \leq 0$, the composition needs to ensure the convexity of the new set of transformed constraints, see \cite[Section~3.2.4]{Boyd-CVXBook}. More details on incorporating general local constraints in the objective can be found in \cite{Boyd-CVXBook,mikael2021cdc}.}. It is assumed that $a_i\neq 0$ for all $i$.

%The proposed solutions need to be distributed, implying that the information available at each agent $i$ includes its own information (for example, its state and local objective) and the data received from agents $j \in \mc{N}_i$  (its direct neighbours). In this work, it is assumed that agents are subject to some model nonlinearities. Moreover, the distributed solution must remain feasible (i.e.,  $\sum_{i=1}^{n} x_i(t) = b$, $\forall t>0$) and be delay-tolerant. Further, it is possible that the communication network changes over time and loses connectivity. The assumptions on the objective function convexity (to include possible penalty terms), the network connectivity, feasibility, and the time-delay model are essential in the problem setup, as discussed next.

\section{Preliminaries} \label{sec_pre}
\subsection{Lipschitz Continuity, Smoothness, and Convexity for Distributed Optimization} \label{sec_lemcon}

Different properties of the objective (or cost) function affect the distributed solution proposed in the literature. The first one is on the smoothness of the objective.
The objective function $f: \mathbb{R} \mapsto  \mathbb{R} $ is called Lipschitz continuous if there is $K_f \in \mathbb{R}$ such that for any $x_1, x_2 \in \mathbb{R}$,
\begin{align}
|f(x_1)-f(x_2)| \leq K_f |x_1-x_2|.	
\end{align}
The smoothness follows the Lipschitz continuity of the function derivative. If the function is Lipschitz and smooth, typical derivative-based (or gradient-based) convergence analysis can be adopted. Otherwise,  the convergence analysis is based on non-smooth models \cite{cortes2008discontinuous,doostmohammadian20211st}.

For a convex function $f:\mathbb{R} \mapsto \mathbb{R}$ and for all $ x_1,x_2 \in \mathbb{R}$ and $\kappa \in (0,1)$, it holds that
\begin{align}
f(\kappa x_1+(1-\kappa)x_2) \leq \kappa f(x_1)+(1-\kappa)f(x_2).
\end{align}
where the strict inequality holds for strictly convex functions. For a smooth strictly convex function one can show that $\partial^2 f(x)> 0$
for $x \in \mathbb{R}$ \cite{bertsekas_lecture}.
%Recalling the Taylor series expansion, the following holds.
%\begin{lemma} \label{lem_strict}
%\cite{boyd2006optimal} Given a continuous strictly-convex function $f(y)$, two points $y_1, y_2$, and $\Delta y =: y_1-y_2$, there exist  $\overline{y} := \kappa y_1 + (1-\kappa)y_2, 0<\kappa<1 $ such that,
%\begin{align}
%	f(y_1) = f(y_2) + \nabla F(y_2)^\top \Delta y +  \frac{1}{2} \Delta y^\top \partial^2 f(\overline{y})\Delta y. \label{eq_taylor}
%\end{align}
%\end{lemma}
%
%\begin{lemma} \label{lem_strict2}
%\cite{bertsekas_lecture}  For cost function ${f}_i:\mathbb{R} \mapsto \mathbb{R}$, assume $2 v < \partial^2 f_i < 2 u $ with $0<v \leq u <\infty$\footnote{The condition $2 v  < \partial^2 f_i(x_i) $ implies that the cost function  ${f}_i$ is strongly convex, see Assumption~\ref{ass_lips_strict} in Section~\ref{sec_rate}.}. Then,  for two points $\mb{x}_1, \mb{x}_2 \in \mathbb{R}$, and ${\Delta x := \mb{x}_1 - \mb{x}_2}$, the following statements hold:
%%	there exist  $\overline{\mb{x}}_i = \kappax_i(k+1) + (1-\kappa)x_i(k), 0<\kappa<1 $ such that,
%\begin{align}
%	F(x_1) &< F(x_2) + \nabla F(x_1) \Delta x + {u} \Delta x \Delta x. \label{eq_taylor1} \\ \nonumber
%	F(x_1) &> F(x_2) + \nabla F(x_1) \Delta x + {v} \Delta x^\top \Delta x. \label{eq_taylor2}
%\end{align}
%\end{lemma}
For convex objective functions, it is proved that the optimal solution $\mb{x}^*=[{x}^*_1;\dots;{x}^*_n]$ to problem \eqref{eq_dra} satisfies the following \cite{bertsekas_lecture,boyd2006optimal},
\begin{align} \label{eq_optimalX}
\nabla {F}(\mb{x}^*) = \varphi^*\mb{1}_n,
\end{align}
with $\mb{1}_n$ representing the vector of all $1$s,
%$\partial f_i \in \mathbb{R} $ as the gradient of the local  function $f_i(\cdot)$,
$\nabla F(\mb{x}^*) = [\partial f_1(\mb{x}^*_1); \dots; \partial f_n(\mb{x}^*_n)] \in \mathbb{R}^{n}$, and $\varphi^* \in \mathbb{R}$. This directly follows from the well-known Karush-Kahn-Tucker (KKT) condition and the method of Lagrange multipliers. Similarly, it can be proved that for the solution to the original problem \eqref{eq_dra0} we have,
\begin{align} \label{eq_optimalZ}
\nabla \widetilde{F}(\mb{z}^*) = \widetilde{\varphi}^* \mb{a}.
\end{align}
Given that the objective function in \eqref{eq_dra0} and \eqref{eq_dra} is strictly convex, the optimal solution is unique.
%Note that, in the presence of the box constraints, the above lemma assumes that $\mb{z}^*$ meets those constraints, i.e., $\underline{z}_i \leq z^*_i \leq \overline{z}_i$ for all $i$.
This is proved by another relevant concept known as the level set analysis. For $F(\mb{x}): \mathbb{R}^{n} \mapsto \mathbb{R} $, one can define the level set $L_\gamma(F)$ for $\gamma \in \mathbb{R}$ is the set $L_\gamma(F)= \{\mb{x} \in \mathbb{R}^{ n}|F(\mb{x})\leq \gamma\}$. Let assume $F(\mb{x})$ is convex; then, all its level sets $L_\gamma(F)$ are also convex $\forall \gamma >0$ \cite{bertsekas_lecture}. For two distinct points $\mb{x}_1$ and $\mb{x}_2$, let $F(\mb{x}_1)>F(\mb{x}_2)$. For two associated level sets $L_{\gamma_1}(F)$ and $L_{\gamma_2}(F)$ with $\gamma_1=F(\mb{x}_1),\gamma_2=F(\mb{x}_2)$ one can prove that \cite{lj_lecture},
\begin{equation} \label{eq_level}
\nabla F(\mb{x}_1)^\top(\mb{x}_2-\mb{x}_1) \geq 0.
\end{equation}
For strictly convex function $F$, the strict inequality in \eqref{eq_level} holds.
It can be proved that for every feasible set $\mc{S}_\mb{b}$ and with a strictly convex function there is a unique solution $\mb{x}^* \in \mc{S}_\mb{b}$  that satisfies Eq. \eqref{eq_optimalX} (similar statement holds for $\mb{z}^* \in \widetilde{\mc{S}}_b$ and Eq. \eqref{eq_optimalZ}) \cite{doostmohammadian20211st}.
This is because the level sets of $F(\mb{x})$ are strictly convex and implies that only one of its level sets say $L_\gamma(\widetilde{F})$, is adjacent to the constraint facet $\mc{S}_\mb{b}$, where the touching only happens at one point, say $\mb{x}^*$. Since the level set is adjacent to the constraint at $\mb{x}^*$, the gradient $\nabla F(\mb{x}^*)$ is orthogonal to the constraint facet $\mc{S}_\mb{b}$ and follows that
\begin{eqnarray}
\nabla F(\mb{x}^*) \in \mbox{span}(\mb{1}_n).
\end{eqnarray}	
Using contradiction, consider $\mb{x}^{*}_1$ and $\mb{x}^{*}_2$ in $\mc{S}_\mb{b}$ satisfying $\nabla F(\mb{x}^{*}_1) = \alpha_1  \mb{1}_n$ and $\nabla F(\mb{x}^{*}_2) = \alpha_2  \mb{1}_n$. This can be interpreted in two ways: (i) we have two points for which $L_\gamma(F)$, $\gamma = F(\mb{x}^{*}_1) = F(\mb{x}^{*}_2)$ touches the constraint $\mc{S}_\mb{b}$ which is an affine set. This contradicts the strict convexity of the level set $L_\gamma(F)$;
(ii) the points are on the border of two level sets $L_{\gamma_1}(F),~\gamma_1 = F(\mb{x}^{*}_1)$ and $L_{\gamma_2}(F),~\gamma_2= F(\mb{x}^{*}_2)$, which are both adjacent to the same affine set $\mc{S}_\mb{b}$. Now, assume $F(\mb{x}^{*}_2)>F(\mb{x}^{*}_1)$, and following \eqref{eq_level} we have,
\begin{equation} \label{eq_proof1}
\nabla F(\mb{x}^{*}_2)^\top(\mb{x}^{*}_1-\mb{x}^{*}_2) > 0.
\end{equation}
However, since both  $\mb{x}^{*}_1$ and $\mb{x}^{*}_2$ belong to $\mc{S}_\mb{b}$  we have $\mb{1}_n^\top (\mb{x}^{*}_1-\mb{x}^{*}_2) = 0$ with the level sets both adjacent to $\mc{S}_\mb{b}$ at both points. This implies that $\nabla F(\mb{x}^{*}_2)^\top(\mb{x}^{*}_1-\mb{x}^{*}_2) = 0$,   which contradicts \eqref{eq_proof1}.
Since both of the mentioned cases (i) and (ii) contradict the properties of the strictly convex level sets, this implies the uniqueness of $\mb{x}^{*} \in \mc{S}_\mb{b}$ satisfying Eq.~\eqref{eq_optimalX} for strictly convex objective functions.

It is typically assumed that the local objective functions ${f}_i(x_i):\mathbb{R} \mapsto \mathbb{R}$, $i \in \{1,\dots,n\}$ satisfy $\partial^2 f_i(x_i) > 0$ in primal-based formulations and in dual-based formulations only convex assumption is made (see \cite{nedic2018distributed,yang2019survey} and references therein)\footnote{Primal and dual formulations are later discussed in details in Section~\ref{sec_sol}.}. Note that the penalty term $ f_i^{\sigma}$ (for every node $i$) is Lipschitz continuous and smooth for $\sigma \in \mathbb{N}_{\geq 2}$ and non-smooth for $\sigma =1$. Therefore, in primal-based formulations (as in the Laplacian-gradient solution) it is common to consider $\sigma \in \mathbb{N}_{\geq 2}$.

Different assumptions on the convexity and smoothness of the objective function are discussed in the literature. Laplacian-gradient-based solutions mainly assume smooth and strictly (or strongly) convex functions. In the case of non-smooth but locally Lipschitz objective functions, the generalized gradient in the following form is used instead \cite{doostmohammadian20211st,cortes2008discontinuous},
\begin{align}
\partial f(x)= \mathrm{co}\{\lim \nabla f(x_i): x_i \rightarrow x, x_i \notin \Omega_f \cup S \},
\end{align}
with $\mathrm{co}$ as the convex hull, $S \subset \mathbb{R}$ as any set of zero Lebesgue measure, and $\Omega_f \in \mathbb{R}$ denoting the non-differentiable set of points. If $f$ is locally Lipschitz at $x$, then $\partial f(\mb{x})$ is nonempty, compact, and convex, and the set-valued map $\partial f:\mathbb{R} \rightarrow \mc{B}\{\mathbb{R}\}$ (with $\mc{B}\{\mathbb{R}\}$ as all subsets of $\mathbb{R}$), $\mb{x} \mapsto \partial f(\mb{x})$, is upper semi-continuous and locally bounded \cite{cortes2008discontinuous}.

\subsection{Algebraic Graph Theory} \label{sec_graph}
The network of nodes/agents\footnote{Throughout this paper the terms nodes or agents, links or edges, and network or graph are used interchangeably. }  (or the multi-agent system) is typically modelled by a (directed) graph $\mc{G}=\{\mc{V},\mc{E}\}$, where the set of links (or edges) $\mc{E}$ denotes the interaction of different nodes/units $\mc{V}=\{1,\dots,n\}$. A directed link $(i,j) \in \mc{E}$ implies data transmission from node $i$ to node $j$. This also defines the neighbouring set of nodes, i.e., $i$ is a neighbour of $j$. In general, two neighbouring sets of nodes are defined, the set of in-neighbours $\mc{N}^-_j=\{i|(i,j)\in \mc{E}\}$ and the set of out-neighbours $\mc{N}^+_j=\{i|(j,i)\in \mc{E}\}$. The graph is associated with a weight matrix $W$ (also referred to as the adjacency matrix). Each link $(j,i) \in \mc{E}$ is associated with the weight $W_{ij}>0$ and the matrix $W :=[W_{ij}] \in \mathbb{R}^{n \times n}_{\geq0}$. The zero-nonzero pattern of $W$ follows the topology of $\mc{G}$.
Define the associated Laplacian matrix
$L=[L_{ij}]$ as,
\begin{align} \label{eq_laplacian}
L_{ij} = \left\{ \begin{array}{ll}
	\sum_{j \in \mc{N}^-_i} W_{ij}, & \text{for}~  i=j,\\
	-W_{ij}, & \text{for}~i\neq j.
\end{array}\right.
\end{align}
Following the definition, $\mb{1}_n^\top L = \mb{0}_n$ and $L$ has (at least) one zero eigenvalue. More eigen-spectrum properties of the Laplacian matrix can be found in \cite{godsil,olfati_rev}.

There are various assumptions on graph connectivity in the literature. The network could be undirected (bidirectional links), directed and balanced, time-varying (switching), or uniformly connected (B-connected). Existing works in the literature also consider unbalanced digraphs, e.g., see \cite{zhang2020distributed}.  Each case is defined below.
\begin{itemize}
\item \textbf{Undirected and Symmetric:} Every link in $\mc{G}$ is bidirectional with symmetric weight matrix, i.e., $W_{ij}=W_{ji}$. This implies a symmetric weight matrix $W$ and symmetric Laplacian $L$.
\item \textbf{Balanced Digraph\footnote{There exist distributed algorithms in the literature to ensure weight-balance or weight-stochasticity over the directed networks, see for example \cite{gharesifard2012distributed,hadjicostis2018distributed}.
	}:} The graph is possibly directed with the link weights satisfying
\begin{align} \label{eq_bal1}
	\mb{1}_n^\top W = W \mb{1}_n,
\end{align}
or
\begin{align} \label{eq_bal2}
	\sum_{j \in \mc{N}^-_i} W_{ij} = \sum_{j \in \mc{N}^+_i} W_{ji},~\forall i\in \mc{V}.
\end{align}
\item \textbf{Stochastic:} A network is weight-stochastic (or row-stochastic) if
\begin{align} \label{eq_stoc1}
	W \mb{1}_n = \mb{1}_n,
\end{align}
and is bi-stochastic (or doubly stochastic) if it is balanced and,
\begin{align} \label{eq_stoc2}
	W \mb{1}_n  = W^\top \mb{1}_n = \mb{1}_n.
\end{align}
\item \textbf{Switching:} In case $\mc{G}(t)=\{\mc{V},\mc{E}(t)\}$ is time-varying (with the same node set $\mc{V}$) or the link weights change over time the network is called of switching topology.
\item \textbf{Uniformly-Connected:} For a time-varying network if the union graph $\mc{G}_B(t)=\{\mc{V},\mc{E}_B(t)\}$ over a time-window $B>0$ is connected  it is called B-connected or uniformly-connected. In other words, there is $B>0$ such that the (edge) union graph (defined as follows) is connected for $t \geq 0$.
\begin{align} \label{eq_uniformly}
	\mc{E}_B(t) = \bigcup_{t}^{t+B}\mc{E}(t),~\mc{G}_B(t) = \bigcup_{t}^{t+B}\mc{G}(t).
\end{align}
\end{itemize}

The network connectivity depends, for example, on the broadcasting power levels and the communication range of the agents or their mobility over time.
%Therefore, if $i \in \mc{N}^-_j(t)$ then $j \in \mc{N}^-_i(t)$ or $j \notin \mc{N}^-_i(t)$ defines the network connectivity.
The uniform-connectivity is the weakest form of network connectivity over time. This implies $\mc{G}(t)$ may lose connectivity at some times while possibly regaining connectivity at some other times. In other words, the connectivity holds over longer time intervals in the case of communication links arbitrarily connecting and disconnecting over the dynamic network. This is the weakest requirement in distributed coordination and consensus literature \cite{nedic2014distributed,ren2005consensus} however it is strong enough to guarantee convergence.
Different network-connectivity assumptions are given in the literature. The works \cite{falsone2020tracking,rikos2021optimal} assume weight-stochastic condition while all-time connectivity assumption is considered in \cite{banjac2019decentralized,cherukuri2015distributed,mikael2021cdc,boyd2006optimal}.
Recall that for switching and stochastic networks it is needed to re-adjust (update) the link weights to ensure weight-stochasticity at all times, see \cite{6426252} for such algorithms.

It is known that the convergence rate of the optimization and scheduling algorithms is tied with the eigen-spectrum of the Laplacian matrix  $L$. Among the eigenvalues, the largest eigenvalue $\lambda_n$ and smallest nonzero eigenvalue $\lambda_2$ play a key role in the convergence rate and the optimization step rate. This is a result of Courant-Fischer
theorem and consensus algorithms \cite{SensNets:Olfati04}. For example,
for a balanced Laplacian matrix $L$ and state vector $\mb{x} \in \mathbb{R}^n$ and  $\overline{\mb{x}} =: \mb{x} - \frac{\mb{1}_n^\top \mb{x}}{n} \mb{1}_n$ (known as dispersion state vector), a common (disagreement) Lyapunov function is used as follows:
\begin{align} \label{eq_laplace1}
\mb{x}^\top L_s \mb{x} &= \overline{\mb{x}}^\top L_s \overline{\mb{x}},
%\mb{x}^\top L \mb{y} = \overline{\mb{x}}^\top L \overline{\mb{y}},
% \\ \label{eq_laplace2}
%  &\mb{x}^\top L \mb{y} <  \mb{y}^\top L^2 \mb{y} ~~\mbox{\mb{if}}~~  \overline{\mb{x}}^\top \overline{\mb{x}} < \overline{\mb{y}}^\top L^2 \overline{\mb{y}}
\\      \label{eq_laplace}
\lambda_2 \|\overline{\mb{x}} \|_2^2 \leq \mb{x}^\top &L_s\mb{x} \leq \lambda_n \|\overline{\mb{x}} \|_2^2,
\end{align}
with  $L_s = \frac{L+L^\top}{2}$. Note that for (undirected) connected graphs with weights equal to $1$, $\lambda_2$ is called the Fiedler value or the Algebraic connectivity  \cite{SensNets:Olfati04,graph_handbook} and is representative of the connectivity of the associated network.
For uniformly-connected network $\mc{G}_B$, $\lambda_n \leq \lambda_{nB}$ and $\lambda_2 \leq \lambda_{2B}$. This implies that link addition may increase the algebraic connectivity \cite{wang2008algebraic,SensNets:Olfati04}. Therefore, given $\mc{G}= \mc{G}_1 \cup \mc{G}_2$, the algebraic connectivity implies that $\lambda_2(\mc{G}) \geq \lambda_2(\mc{G}_1),\lambda_2(\mc{G}) \geq \lambda_2(\mc{G}_2)$. Therefore, for uniform-connectivity, we have $\lambda_2(\mc{G}_B(t)) \geq \lambda_2(\mc{G}(t))$. One way to prove this is by the fact that $\lambda_2(\mc{G}) \geq \frac{1}{nd_g}$ (with $d_g$ as the network diameter) \cite[p. 571]{graph_handbook}. Note that the eigen-spectrum of the network is key in the convergence rate \cite{olfati_rev} and the upper bound on the step-rate of the optimization algorithms \cite{ojcsys}.

\subsection{Consensus and Coordination Algorithms}
In control theory, consensus in distributed systems implies achieving agreement (or consensus) on a certain quantity/property among a group of interconnected agents or nodes.
The goal of consensus algorithms
are to synchronize the states or behaviours of these agents so that they reach a common decision or converge to a shared value over a network \cite{olfati_rev,ren2005consensus,renetal05}.

The intuition of a typical consensus algorithm is the following. Each agent represents a node or component that interacts with its neighbours by exchanging information.
Consensus algorithms enable agents to communicate and update their states based on the information received from neighbouring agents. The iterative process continues until a consensus or agreement is reached, typically by adjusting the states of the agents to minimize the differences between their values.

Consensus algorithms achieve convergence when the states of all agents become sufficiently close/equal to each other. Convergence is typically ensured by carefully choosing the weights and the communication topology among the agents \cite{olfati_rev}. Different weight assignment strategies, such as average consensus \cite{olfati_rev,ren2005consensus}, consensus with leader \cite{cao2015leader,tanner02}, or consensus with time-varying weights \cite{xiao2008asynchronous}, can be used based on specific system requirements and objectives.

Consensus algorithms find applications in networked control systems \cite{kia2019tutorial}, e.g., coverage control \cite{msc09,MiadCons,cortes2004coverage,bullo2009distributed,telsang2022decentralized}, opinion dynamics \cite{bullo-opinion,jstsp14}, rendezvous algorithms \cite{cortes2006robust}, flocking and formation algorithms \cite{Jadbabaie_flocking,olfati2002distributed}, target tracking \cite{flock,taes}, FDI \cite{tcns_fdi,ijc_fdi}, attack detection \cite{giraldo2018survey,deghat2019detection}, and estimation \cite{jstsp,kar2013consensus,nuno-suff.ness}.
More recent applications are in machine learning and optimization \cite{xin2020decentralized,dsvm}. These algorithms enable agents to coordinate their actions, reach agreements, and collectively solve problems in a distributed manner.
In general, consensus algorithms are crucial for achieving cooperation, coordination, and synchronization among agents in distributed systems.

There are various consensus algorithms used in control theory, with the most well-known being the consensus protocol based on linear consensus dynamics \cite{sherayas:08,olfati_rev,ren2005consensus}. In this protocol, the agents update their states by incorporating a weighted average of their own state and the states of their neighbours, and the weights determine the influence of each neighbour on the agent's state update. The state update in discrete time is the following:
\begin{align}
\mb{x}(k+1) = W \mb{x}(k),
\end{align}
with $\mb{x}(k)$ being the states of the agents at time step $k$,  and $W$ being a stochastic matrix (or more accurately  stochastic-indecomposable-aperiodic (SIA) \cite{ren2005consensus}). In continuous-time, the state update can be expressed as:
\begin{align}
\dot{\mb{x}}(t) =- L \mb{x}(t),
\end{align}
with $L$ being the Laplacian matrix. It is known that the convergence rate of these consensus protocols and other consensus-based distributed algorithms tightly depends on the algebraic graph properties and spectrum analysis of the associated stochastic matrix $W$ or the Laplacian matrix $L$ (discussed in Section~\ref{sec_graph}).
In terms of convergence scheme, other than the asymptotic convergence in linear models, finite-time \cite{taes2020finite,feng2017finite,hu2018distributed,lin2016distributed,bhat2000finite,Scientia2011}, fixed-time \cite{Garg_fixed,shang2017,ning2017distributed,firouzbahrami2022cooperative}, and prescribed-time \cite{wang2018prescribed,ji2023initialization} models are proposed to guarantee convergence over a finite bounded time-interval in contrast to asymptotic convergence. These algorithms mainly leverage non-Lipschitz sign-based dynamics to accelerate the agreement and convergence.

\section{Distributed Resource Allocation Algorithms} \label{sec_sol}
The existing literature on distributed optimization includes both \textit{constrained} and \textit{unconstrained} scenarios. These two scenarios are directly related as described in \cite{zhang2020distributed} (and discussed later in Section~\ref{sec_dual}).
The literature on distributed optimization can be classified in terms of being general or problem-specific
based on the nature of the optimization problem being addressed. Most works in the literature assume ideal linear models to describe the data exchange among agents in the network, and consider no problem-specific (nonlinear) constraints on the agents' actuation dynamics.
There are also some works in the literature that address problem-specific constraints in their proposed solutions and other kinds of constraints in their models. In the following subsections, we discuss and analyze both kinds of distributed optimization algorithms.

\subsection{General Algorithms}
In the context of distributed optimization, linear models are based on techniques like consensus optimization. These include both unconstrained problems  \cite{kia2015distributed,Uribe2020,shreyas2020byzantine,khan2020optimization,xin2019frost,qureshi2020s,agarwal2012delay,al2020gradient},
and problems subject to state constraints \cite{dsvm,boyd2006optimal,gharesifard2013distributed,shames2011accelerated,doan2017scl,doan2017ccta,nedic2018improved,Turan2021,Uribe19,sayed2019proximal,hamedani2017multi,aybat2016distributed,yi2016initialization,yang2013consensus,wang2018distributed,cherukuri2015distributed,jiang2021distributed,lakshmanan2008decentralized}. These two are related via the mirror relation between distributed optimization and dual formulation of the DRA described in Section~\ref{sec_dual}.
Other approaches in the literature include, diffusion-based \cite{sayed2019proximal}, event-triggered \cite{kia2015distributed}, primal-dual  \cite{Uribe2020,Turan2021,hamedani2017multi}, Lagrangian/ADMM-based \cite{doan2017ccta,aybat2016distributed,jiang2021distributed}, gradient push-sum \cite{yang2013consensus}, push-pull (or AB type algorithm) \cite{xin2019frost}, attack-resilient \cite{shreyas2020byzantine,Turan2021}, and momentum-based (or accelerated) solutions \cite{Uribe19,shames2011accelerated}. The solutions and their convergence analysis are given in both continuous-time \cite{dsvm,kia2015distributed,gharesifard2013distributed,Uribe19,yi2016initialization,ning2017distributed,wang2018distributed,cherukuri2015distributed} or discrete-time dynamics  \cite{sayed2019proximal,xin2019frost,khan2020optimization,qureshi2020s,doan2017scl,doan2017ccta,nedic2018improved,Uribe2020,shreyas2020byzantine,hamedani2017multi,aybat2016distributed,yang2013consensus,jiang2021distributed,lakshmanan2008decentralized}. Other than the \textit{resource allocation problem} in which the coupling constraint is on resource-demand feasibility, some works in the literature address the distributed optimization subject to consensus-constraint \cite{dsvm,sayed2019proximal,wang2018distributed,jiang2021distributed}.
On the other hand, coupling constraints on the sum of resources are considered in the DRA literature, e.g., see the linear preliminary works in \cite{doan2017scl,doan2017ccta,nedic2018improved,Turan2021,Uribe19,cherukuri2015distributed,lakshmanan2008decentralized}. However, note that the aforementioned literature assumes no extra constraints on the agents' dynamics such as nonlinearities or time-delays in the communication links and/or agents' actuation.
The existing general DRA algorithms are categorized into primal-based, dual-based, and ADMM-based solutions which are studied in the next subsections in detail.

\subsubsection{Laplacian Gradient Solutions to the Primal-Formulation} \label{sec_lg}
Laplacian gradient solutions apply the concept of Laplacian matrix $L$ to coordinate and optimize the resource allocation across multiple agents or nodes in a network. Each agent $i$ updates its local variables based on the gradients of the primal objective functions (its own and its neighbours). The local variable update is given by:
\begin{align}\label{eq_sol}
	x_i (k+1&) = x_i(k) -\alpha \sum_{j \in \mc{N}^-_i} W_{ij} (\partial_{x_i}f_i(k) - \partial_{x_j}f_j(k)) ,
\end{align}
with  $\partial_{x_i}f_i$ as the derivative of $f_i$ with respect to $x_i$, and $\alpha$ as the step-size. In case the function is nonsmooth, the generalized gradient is used instead in Eq.~\eqref{eq_sol}. With a slight abuse of notation, one can rewrite the above solution in compact Laplacian-gradient (LG) form:
\begin{align} \label{eq_sol_L}
	\mb{x}(k+1) = \mb{x}(k) -
	\alpha L \nabla F(k) ,
\end{align}
where the bound on step-size $\alpha$ to ensure convergence can be defined based on algebraic properties of the associated Laplacian matrix (the eigen-spectrum properties in Section~\ref{sec_graph}), the convexity properties of the objective function; see \cite{boyd2006optimal,ojcsys} for more details.
The equivalent continuous-time version of~\eqref{eq_sol_L} can be written in the following form:
\begin{align} \label{eq_sol_L2}
	\dot{\mb{x}} = -
	\alpha L \nabla F .
\end{align}
For strictly convex objective functions and initializing from a state that satisfies resource-demand constraint (a feasible initialization) there is only one unique optimal convergence point to the above dynamics that matches the solution of the standard DRA problem~\eqref{eq_dra} in the form~\eqref{eq_optimalX}. This is proved by level set analysis in Section~\ref{sec_lemcon} (see more details in \cite{fast,doostmohammadian20211st}).
The main feature of this approach is that, for feasible initialization, resource-demand constraint $\sum_{i=1}^n x_i = b$ holds at all times and all iterations of the solution dynamics. This is because we have
\begin{align}
	\mb{1}_n^\top \mb{x}(k+1) = \mb{1}_n^\top \mb{x}(k) -
	\alpha \mb{1}_n^\top L \nabla F(k) = \mb{1}_n^\top \mb{x}(k) ,
\end{align}
where the last equality follows from the fact that $\mb{1}_n^\top$ is the left eigenvector associated with the zero eigenvalue of the Laplacian matrix $L$.
This is referred to as all-time feasibility (or anytime feasibility) and is discussed later in Section~\ref{sec_feas}.
For the given dynamics \eqref{eq_sol} or \eqref{eq_sol_L2} one can see that the equilibrium point $\mb{x}^*$ is only in the form $\nabla F(\mb{x}^*) \in \mbox{span}(\mb{1}_n)$ for  which $\dot{\mb{x}} = \mb{0}$ or $ \mb{x}(k+1) = \mb{x}(k)$. One can prove by contradiction that there is no other equilibrium for the dynamics \eqref{eq_sol} or \eqref{eq_sol_L2} with $\nabla F(\mb{x}^*) \notin \mbox{span}(\mb{1}_n)$. Suppose by contradiction that the equilibrium point $\mb{x}^*$ satisfies $	\partial_{x_i} f_i(\mb{x}^*_i) \neq  \partial_{x_j} f_j(\mb{x}^*_j)$
for (at least) two nodes $i,j$. Let $\alpha = \argmax_{i\in \{1,\dots,n\}}  \partial_{x_i} f_i(\mb{x}^*_i)$ and $\beta = \argmin_{i \in \{1,\dots,n\}}  \partial_{x_i} f_i(\mb{x}^*_i)$ over the multi-agent system.	
Then, over the network of agents   $\mc{G}(t)$ (or its uniformly connected union $\mc{G}_B(t)$) there is (at least) one path between agents $\alpha$ and $\beta$ or vice versa. Over such a path, there exists two agents $\overline{\alpha}$ and $\overline{\beta}$ for which
$\partial_{x_i} f_{\overline{\alpha}}(\mb{x}_{\overline{\alpha}}^*)  \geq \partial_{x_i} f_{\mc{N}_{\overline{\alpha}}}(\mb{x}^*_{\mc{N}_{\overline{\alpha}}}) , \partial_{x_i} f_{\overline{\beta}}(\mb{x}^*_{\overline{\beta}})  \leq \partial_{x_i} f_{\mc{N}_{\overline{\beta}}}(\mb{x}^*_{\mc{N}_{\overline{\beta}}}) $
with $\mc{N}_{\overline{\alpha}}$ and $\mc{N}_{\overline{\beta}}$  respectively as the set of neighbours of $\overline{\alpha}$ and $\overline{\beta}$ with the strict inequality for (at least) one neighbour in $\mc{N}_{\overline{\alpha}}$ and $\mc{N}_{\overline{\beta}}$. Therefore, if the network is always connected or uniformly connected over time we get $\dot{\mb{x}}^*_{\overline{\alpha}} < 0$ and   $\dot{\mb{x}}^*_{\overline{\beta}} > 0$ or in discrete-time $\mb{x}^*_{\overline{\alpha}}(k+1) < \mb{x}^*_{\overline{\alpha}}(k)$ and   $\mb{x}^*_{\overline{\beta}}(k+1) > \mb{x}^*_{\overline{\beta}}(k)$. These contradict the definition of equilibrium for $\mb{x}^*$ and imply that for the LG-type dynamics in the form \eqref{eq_sol} or \eqref{eq_sol_L2} one can easily prove that the equilibrium $\nabla F(\mb{x}^*) \in \mbox{span}(\mb{1}_n)$ over uniformly-connected multi-agent networks.

\subsubsection{Dual-Problem-based Gradient Tracking Solutions} \label{sec_dual}
Dual-problem-based gradient tracking solutions use the concept of dual variables and gradient tracking to satisfy the equality constraints while optimizing the objective function \cite{zhang2020distributed}. Here we explain dual-problem-based gradient tracking solutions for equality-constraint DRA. The existing methods exploit the
relationship between distributed optimization and DRA \cite{zhang2020distributed,xu2018dual,nedic2018improved,zhu2019distributed,xu2017distributed,alghunaim2021dual,li2022implicit,wu2022distributed} and the idea of consensus-based small-gain theorem \cite{zholbaryssov2022fast}. The dual problem is formulated by introducing dual variables corresponding to the equality constraints. Let $\lambda_i$ be the dual variable associated with the (more general) equality constraint $A_i x_i = b_i$. The dual problem aims to maximize the dual function with respect to the dual variables $\lambda_i$ and is defined as:
\begin{align} \label{eq_dual}
	\max_{\lambda_i} \Phi(\lambda) = \min_x	L_\rho (x,\lambda),
\end{align}
where $L(x,\lambda)$ is the Lagrangian function defined as:
\begin{align} \label{eq_L_dual}
	L(x,\lambda) = F(x) + \sum_{i=1}^n \lambda_i^\top (A_ix_i-b_i).
\end{align}
The key idea for solving~\eqref{eq_dual} is to track the gradients of the Lagrangian function with respect to the dual variables. One solution is that each agent
$i$ updates its dual variable based on the gradients obtained from the Lagrangian function as:
\begin{align} \label{eq_L_lambda}
	\lambda_i(k+1) = \lambda_i(k) + \alpha(k) \nabla_{\lambda_i} L(x(k),\lambda(k)) ,
\end{align}
with $\alpha(k)$ as the step-size (or the learning rate) at iteration $k$. After updating the dual variables, each agent
$i$ solves its local optimization problem to update the local variables
$x_i$ while keeping the dual variables fixed. This can be done using standard optimization techniques to minimize the local objective function subject to the local equality constraint. A consensus step is then performed to improve agreement and convergence among the local variables \cite{zhang2020distributed,alghunaim2021dual}. Specifically, the agents exchange information and update their local variables to improve consensus. This step ensures that the local variable solutions satisfy the global equality constraints. The algorithm continues executing with these three steps until a convergence criterion is met, which may rely on various criteria such as the change in the objective function, the primal and dual residuals, or the number of iterations. There are also primal-dual-based DRA dynamics to optimize \textit{non-smooth} objective functions \cite{huang2024distributed,10176349}.

\subsubsection{ADMM-based Solutions} \label{sec_admm}
The Alternating-Direction-Method-of-Multipliers (ADMM) is an optimization technique that is particularly useful for solving problems with separable objective functions or constraints. It is an iterative method that combines the advantages of decomposition, proximal methods, and dual decomposition. This method is shown to be very effective
for solving large-scale distributed optimization and learning models \cite{boyd2011distributed,iutzeler2015explicit}. Classic ADMM methods are based on a semi-centralized method, known as Parallel-ADMM \cite{bertsekas2015parallel}. Recently, many variations of the distributed version of this method are proposed in the literature \cite{wei2012distributed,chang2016proximal,wei_me_cdc,dtac,falsone2018distributed,aybat2019distributed,makhdoumi2017convergence}. There are many works devoted to dual consensus ADMM \cite{chang2016proximal,chang2014multi,banjac2019decentralized,jian2019distributed} where the consensus ADMM cannot be applied on the primal-formulation but it is adopted on the dual formulation. Some other literature is focused on Jacobi-like ADMM \cite{deng2017parallel}, and proximal ADMM \cite{yang2022proximal}.

Here, we present a common distributed ADMM model for resource allocation (see \cite{yang2022proximal} for details). The local constraint  at each agent $i$ is denoted as $g_i(x_i)$
and captures the local resource allocation requirements or limitations. In the equality-constraint resource allocation formulation, $g_i(x_i)$ is in the form $A_ix_i=b_i$ with the coupling constraint $\sum_i b_i = b$. The ADMM solution is defined over the optimally dual formulation of the original problem based on augmented Lagrangian formulation. The augmented Lagrangian function for ADMM is defined as follows:
\begin{align} \label{eq_Ladmm}
	L_\rho (x,\lambda) = F(x) + \sum_{i=1}^n \lambda_i^\top (A_ix_i-b_i)  + \frac{\rho}{2} \sum_{i=1}^n \|A_ix_i-b_i\|^2.
\end{align}
Here, $\lambda_i$ is the dual variable associated with the agent
$i$, and $\rho$ is a penalty parameter. To solve the problem, the agents perform iterative updates to optimize the local variables and the dual variables. The steps are as follows:
\begin{enumerate} [(i)]
	\item  Local variable update: Each agent
	$i$ solves its local optimization subproblem by minimizing the augmented Lagrangian function with respect to its local variables $x_i$
	while keeping the dual variables fixed. This minimization is defined as:
	\begin{align} \label{eq_Ladmm_x}
		x_i(k+1) = \argmin_{x_i} L_\rho (x,\lambda).
	\end{align}
	This can be achieved using standard optimization techniques, such as gradient-based methods or closed-form solutions depending on the problem structure.
	\item Dual variable update: Each agent
	$i$ updates its dual variable $\lambda_i$
	based on the current local variable solution and the global dual variables:
	\begin{align} \label{eq_Ladmm_lambda}
		\lambda_i(k+1) = \lambda_i(k) + \rho( A_ix_i(k+1)-b_i).
	\end{align}
	\item Consensus update:
	A consensus step is performed to achieve agreement and convergence among the agents. The local variable updates and dual variable updates naturally along with the consensus update ensure consistency between the agents. Note that the primal and dual updates of ADMM both violate the coupling constraint and, therefore, the use of dynamic average consensus algorithms to track
	the feasibility-constraint violation distributedly is adopted.
\end{enumerate}
ADMM iterates until the convergence criterion is met.
%This criterion can be based on the change in the objective function, the primal and dual residuals, or the number of iterations.
The key idea behind ADMM is the alternating updates of $x$ variables while maintaining consistency with the dual variable $\lambda$. The algorithm iteratively solves subproblems that are relatively easier to solve than the original problem. The augmented Lagrangian terms help enforce the constraints and balance the primal-dual convergence.
ADMM is particularly effective when the objective function and constraints can be decomposed into separable parts. This allows for parallel or distributed computation, as different nodes or processors can independently optimize their local variables $x_i$ and communicate their results to update the dual variable $\lambda$.
%The ADMM-based approach for equality-constraint resource allocation aims to find a solution that minimizes the global objective while satisfying the equality constraints for each agent.
By iteratively updating the local variables and the dual variables, ADMM enables agents to coordinate their resource allocations in a distributed manner while respecting the constraints imposed by the problem.
It is important to note that the specific mathematical expressions and algorithms may vary depending on the problem formulation and the specific constraints and objectives of the resource allocation problem at hand. The formulations described here provide a general framework for understanding how ADMM can be applied to equality-constraint resource allocation. Other than resource allocation, ADMM has been successfully applied in various fields, including machine learning, signal processing, image reconstruction, and network optimization. Compared to the gradient-based approaches, ADMM has both advantages and disadvantages. Specifically, ADMM is well-suited for problems that can be decomposed into smaller, easier-to-solve subproblems. In these cases, compared to gradient-based methods, ADMM can be faster, leading to more efficient parallelization \cite{boyd2011distributed}.

\subsection{Problem-Specific Solutions}
Most real-world applications consider nonlinear constraints.
Linear models, although useful in some aspects, are limited in their ability to capture complex dynamics and interactions. By studying problem-specific models, we can develop more realistic and accurate representations of the existing distributed resource allocation systems. {Nonlinear} constraints exist in both node dynamics and data-exchange among the agents. For example, in AGC the node dynamics are constrained with the so-called RRLs that represent saturation-type nonlinear models on the rate of change on the node states \cite{doostmohammadian20211st}. On the other hand, quantization \cite{themis2021cpu} and clipping exist in the communication systems that exchange information among the nodes/agents. Additionally, it is important to note that some nonlinear models add useful properties to the system, for example, faster convergence and finite/fixed-time solutions \cite{fast}.
Many existing solutions to distributed algorithms are mainly limited to finite-time  \cite{taes2020finite,chen2016distributed} or fixed-time \cite{ning2017distributed,parsegov2013fixed,fast,garg2019fixed2} convergence models. Such sign-based dynamics are also prevalent in consensus literature  \cite{taes2020finite,wei2017consensus,stankovic2019robust} which also allow for robust and noise-resilient design. Other works also develop realistic models considering the effect of actuation saturation or clipping in the data channels. Some existing works study data quantization  \cite{nedic2008distributed,rikos2020distributed,rikos_quant,wei2018nonlinear,nekouei2016convergence,nekouei2016performance,Hadjicostis_run:sum} and some study saturation and clipping \cite{wei2018nonlinear,liu2020global}. Such nonlinear modellings make distributed (both constrained and unconstrained) optimization challenging in terms of computation, accuracy, feasibility, optimality, and convergence. Other than these works that consider only a specific problem, some works focus on a \textit{general} model to account for all types of model constraints on the agents' dynamics. These works may include, e.g., a combination of the aforementioned nonlinear models \cite{vtc,doostmohammadian20211st,dsvm2}.
An example problem-specific DRA solution is given below:
\begin{align}\label{eq_sol_L_nonlin}
	x_i (k+1) = x_i(k) -\alpha \sum_{j \in \mc{N}^-_i} W_{ij} g_n\Big(g_l(\partial_{x_i}f_i(k)) - g_l(\partial_{x_j}f_j(k))\Big),
\end{align}
where $g_n(\cdot)$ denotes possible nonlinearity on the node dynamics and $g_l(\cdot)$ denotes possible nonlinearity on the links. A composition of nonlinearities is considered in the above model. For converging to the optimal value, these nonlinearities must follow some specific properties \cite{doostmohammadian20211st,fast}. The convergence is proved under sign-preserving odd nonlinear mappings for both $g_n(\cdot)$ and $g_l(\cdot)$. An example is when the nonlinear mapping follows upper and lower sector-bound conditions, for example, logarithmic quantization \cite{ojcsys}. The feasibility of the solution holds for undirected networks \cite{vtc,doostmohammadian20211st}. In case, there is no node nonlinearity (i.e., $g_n(x)=x$) the solution converges to the optimal allocation over weight-balanced networks \cite{fast,ojcsys}. These nonlinear solutions might be designed to address data quantization, to handle saturation/clipping, to converge in predefined time, and to address resiliency and robustness to noise. Note that it is not easy to address these nonlinearities with the existing dual-formulation-based solutions, e.g., the ADMM method.

\subsection{Convergence Rate}
The convergence rate refers to the speed (number of iterations) at which a distributed optimization algorithm reaches a state that is sufficiently close to the optimal solution.
Fast convergence is important in various applications, e.g., in time-critical applications where a quick response is required, or in scenarios with limited computational resources, as it reduces the overall computation time and resource consumption.
For example, fast convergence is crucial in EDP and AGC over the power grid since the optimization problem needs to be terminated over finite time-intervals (e.g., the applications in Section~\ref{sec_edp} and \ref{sec_agc}).
Moreover, the convergence rate impacts the scalability of the DRA algorithms as they are commonly employed over large-scale systems.

The convergence rate of existing distributed optimization problems and DRA may differ based on the adopted techniques.
Most DRA literature claim linear convergence for their algorithms. In this case, the residual (i.e., the difference between the current state and the optimal state) decreases geometrically (linear in log scale) with respect to the number of iterations \cite{xin2018linear}.
In most existing works, linear convergence is considered to be desirable and indicates a sufficiently-fast convergence rate, e.g., see  \cite{ling2013decentralized,nedic2017achieving,chang2014multi,doan2017scl,scaman2017optimal,olshevsky2015linear,nedic2017geometrically,stich2019local,haddadpour2019trading,nedic2018improved}. Some of the existing algorithms establish linear convergence for particular quadratic-form objective functions; for example, consensus-based scenarios for CPU scheduling \cite{rikos2021optimal} or economic dispatch \cite{Kar6345156}. However, in most cases, the objective is non-quadratic, e.g., because of extra penalty terms to address the box constraints. For non-quadratic objectives, some ADMM-based solutions prove an ergodic rate of convergence for both the residual and feasibility violation \cite{aybat2016distributed,aybat2016primal}. Some optimization papers propose algorithms with superlinear (faster than linear) convergence rates \cite{beck20141,zargham2013accelerated}.
Additionally, different algorithms and techniques for unconstrained decentralized optimization can be used to improve the convergence rate, such as acceleration methods \cite{scaman2017optimal,shames2011accelerated,xin2019distributed}, adaptive learning rates \cite{dauphin2015equilibrated,sayedchen11,chen2011distributed,liuvariance}, or variance reduction techniques \cite{xin2020decentralized,li2020communication,ghadikolaei2020communication,qureshi2021push}.

In general, for problems subject to feasibility constraints the asymptotic convergence should be fast enough to result in near-optimal solutions before the termination time of the algorithm. For all-time feasible but asymptotic solutions (i.e., the algorithms in Section~\ref{sec_lg}), as the algorithm evolves over time the solution gets closer to the optimal value while satisfying the feasibility constraint at all times.
On the other hand, for dual-based and ADMM solutions (i.e., the algorithms in Sections~\ref{sec_dual} and~\ref{sec_admm}) the convergence is ensured to be fast enough to reach all-time feasibility asymptotically as fast as possible. In general, dual-based and ADMM solutions are known to be faster than primal-based solutions in terms of number of iterations for convergence, however, they are computationally more complex and each iteration is more time-consuming.
Some works in the literature propose strategies that converge in finite-time \cite{wang2020distributed,chen2016distributed,li2020distributed2,doostmohammadian20211st}, fixed-time \cite{dai2022consensus,song2021fixed,chen2020initialization,shi2022distributed,li2020distributed3,fast}, or prescribed-time (predefined-time) \cite{guo2022distributed,9609591,lin2020predefined}. Recall that finite-time convergence implies reaching the optimal state in finite time-interval depending on the initial values, while fixed-time convergence does not depend on the initial states. The prescribed-time algorithms, on the other hand, converge in a predefined time-interval irrespective of the initial states. These algorithms are mainly based on non-Lipschitz sign-based mapping, and in discrete-time, may result in unwanted oscillations  (\textit{chattering} phenomena) around the equilibrium.

\section{Main Features and Properties} \label{sec_feature}
In this section, we discuss different key features of the existing DRA solutions in terms of their assumptions on feasibility constraints, network connectivity, delay-tolerance, and packet-drops.
%There are few works in the literature that can address model nonlinearity while relaxing network connectivity, handling latency, and anytime feasibility altogether \cite{vtc,doostmohammadian20211st}.

\subsection{Anytime Feasibility} \label{sec_feas}
In DRA anytime feasibility implies that resource-demand balance holds during all iterations. In other words, as the optimization algorithm evolves in time, the coupling constraint on the states is always satisfied \cite{mikael2021cdc,cherukuri2015distributed}. In real-world applications, by maintaining the resource-demand balance, the system can better meet the quality of service (QoS) guarantees, ensuring that the allocated resources are sufficient to satisfy the desired demand levels. This is of importance in, for example, economic dispatch in power networks or automatic generation control. In this application, this ensures the produced power meets the power load demand at all times. Violating the resource-demand balance may cause service disruption, power delivery issues, and even breakdown in the power grid \cite{mikael2021cdc,cherukuri2015distributed}. So it is crucial to preserve this balance at all times.
Ensuring feasibility in network structures often relies on the crucial feature of weight-balancing.
The majority of LG-based solutions presume the presence of weight-balanced digraphs or undirected graphs to maintain feasibility \cite{cherukuri2015distributed,ojcsys}.

As detailed in Section~\ref{sec_lg}, LG-based solutions hold all-time resource-demand feasibility (the coupling-constraint) for feasible initialization. This indicates that at any termination time of these algorithms, the resource-demand feasibility holds and there would be no constraint violation \cite{fast,ojcsys,doostmohammadian20211st,cherukuri2015distributed}. In contrast, the feasibility in most optimal dual-formulation-based solutions (such as in Section~\ref{sec_dual} and~\ref{sec_admm}) is asymptotic and, therefore, the convergence must be fast enough to reach (almost) feasibility within the algorithm running time \cite{aybat2016distributed,nedic2018improved,wang2019distributed}. In general, anytime feasibility does not hold in the ADMM-based literature, but some works claim to reach feasibility sufficiently fast during the run-time of the algorithm \cite{banjac2019decentralized,falsone2020tracking}. Dynamic average consensus has been used to track the feasibility violation in a distributed way \cite{li2022implicit}. Feasibility-violation tracking is studied in dual-formulation-based gradient-tracking over unbalanced network \cite{zhang2020distributed} and in dual-based ADMM-tracking \cite{kia2017distributed}.
Note that for all-time feasibility, the agents need to take initial values which are feasible, i.e., in the beginning, the sum of resources must meet the demand. There are some algorithms in the literature to set the feasible initialization of states, for example, see the algorithm in \cite{cherukuri2015distributed}. On the other hand, initialization-free solutions are proposed in the literature \cite{lin2021distributed,yi2016initialization,ji2023initialization,chen2020initialization}; these works do not need specific feasible initialization to converge, but they converge to the feasible solution over time.

\subsection{Network Reliability}
In a distributed optimization setting, multiple nodes or entities collaborate to solve a complex optimization problem by sharing information, exchanging messages, and coordinating their actions over a network. The network acts as the communication infrastructure that enables these interactions. In distributed optimization algorithms, the network is typically modelled as an undirected or directed graph with nodes representing the agents and links as their interactions (message-passing or communication). The properties of the network infrastructure are modelled over the graph topology; for example, packet drop is modelled as link failure \cite{icrom} or failed agents as node removal \cite{li2016consensus}.
Additionally, in some optimization literature, the knowledge of the network topology is local \cite{sundaram2016secure}, i.e., each agent only knows its neighbouring topology and not the entire network.

Distributed optimization algorithms often assume the presence of all nodes throughout the optimization process. However, in real-world scenarios, nodes can fail due to hardware failures, software crashes, or network partitions. When a node fails, it disrupts communication and coordination among other nodes, potentially impacting the overall optimization performance. Robustness to node failures becomes crucial to ensure the algorithm can continue making progress even in the presence of such failures \cite{chen2011distributed,sayedchen11}. On the other hand, some works study optimization under the arrival/addition of new nodes \cite{falsone2022sensitivity}.
Some works are devoted to studying optimality and convergence in the presence of deception attacks \cite{fu2021resilient,shao2020distributed}, denial-of-service (DoS) attacks \cite{sardana2009auto}, byzantine man-in-the-middle attacks \cite{turan2020resilient,xu2022resilient,wang2022byzantine,XU2022}, and malicious agents \cite{yemini2022resilience}. In these works the network links or the nodes could be corrupted and under attack. On the other hand, some works propose privacy-preserving strategies for optimal DRA to avoid some of these attacks \cite{beaude2020privacy,angel2020private,lin2020distributed,chen2021distributed,jacquot2019privacy}. In these works the agents avoid to reveal their private information on the constraints, gradients, and
individual solution profile to a third party. Subsequent subsections will delve into the examination of network reliability concerns, encompassing connectivity, latency, and packet drops.

\subsubsection{Networks Subject to Connectivity Considerations}
Distributed optimization algorithms typically rely on iterative processes to reach an optimal solution. The convergence speed depends on how quickly nodes can exchange information and update their local models or variables. If the network experiences frequent disruptions or high latency, the convergence speed may be significantly affected, leading to slower optimization and potentially suboptimal results. Note that in many applications, the network might be dynamic (time-varying or switching). This is, for example, due to agents' mobility or change in their broadcasting range. At some iterations, the network may even get disconnected and regain connectivity over time. Therefore, in many works the convergence is proved over uniformly-connected networks \cite{doostmohammadian20211st,fast,ojcsys,nedic2009distributed}, in contrast to all-time connectivity in many other works \cite{mikael2021cdc,rikos2021optimal,xin2018linear,xin2019distributed}. The uniform connectivity is common in mobile sensor network applications, where some links may connect and disconnect as the mobile sensors (agents or robots) move into and out of broadcast range of one another. The connectivity might be lost at some times due to link failures while preserving uniform connectivity over some finite time intervals. See the mathematical modelling in Section~\ref{sec_graph} for details. In the existing literature, most dual-formulation-based and ADMM-based solutions in Sections~\ref{sec_dual} and~\ref{sec_admm} assume all-time network connectivity. On the other hand, LG-based solutions defined on the primal formulation in Section~\ref{sec_lg} prove convergence over uniformly-connected networks \cite{ojcsys,derm,fast,doostmohammadian20211st,scl}, which is a privilege of these works.

\subsubsection{Networks subject to Time-Delays}
In a distributed setting, nodes often need to exchange information, such as optimization variables, gradients, or updates. If the network experiences delays, it can significantly impact the communication between nodes. Increased communication delays can hinder the convergence of the optimization algorithm and slow down the overall coordination process. The effect of communication delays in coordination and consensus literature is vastly studied, see for example  \cite{SensNets:Olfati04,seuret2008consensus,Themis_delay,acc22delay} and references therein. One trivial approach to handle time delays is to update over a longer time scale after receiving all delayed information \cite{Themis_delay}. However, a more challenging and desirable approach is to update at the same time-scale via all the delayed and non-delayed received information \cite{SensNets:Olfati04,seuret2008consensus,acc22delay}. On the other hand, asynchronous data transmission over the links is considered.
This is the case where the receiver and the transmitter clocks are not necessarily synchronized and are independent. Therefore, distributed algorithms are needed to synchronize the clocks over the communication network to prevent issues raised by asynchronous cases \cite{schenato2011average,schenato2007distributed}.  In the distributed optimization setting latency may cause the solution to diverge. Few works address homogeneous delays or asynchronous data exchange \cite{wang2019distributed,zhu2019distributed}, and some other works discuss how latency affects unconstrained or consensus-constraint optimization \cite{agarwal2012delay,al2020gradient,wang2018distributed,wang2018distributed,themis2011asynchronous}. Note that latency may result in feasibility gap (violation of the coupling-constraint) \cite{wang2019distributed} while LG-based delay-tolerant solutions lead to no feasibility gap over networks, see \cite{ojcsys,chen2017delay}.

Typically, addressing communication time delays is more convenient in LG-type solutions. Consider a network of agents where each agent's available information consists of its own data without delay and all data possibly delayed from its neighbours up to that point. Generally, it is assumed that the delays are upper bounded, for example by $\overline{\tau}$ steps, to prevent packet drops over the links.
The communicated packets over the multi-agent network are time-stamped and every node $i$ knows the time at which agent $j \in \mc{N}_i^-$ sent its data over the link $(j,i)$. To make the mathematical analysis more convenient, define an  indicator function $\mc{I}_{k,ij}$ capturing the delay $\tau_{ij}(k) \leq \overline{\tau}$ on the link $(j,i)$,
\begin{align}
	\mc{I}_{k,ij}(\tau) = \left\{ \begin{array}{ll}
		1, & \text{if}~  \tau_{ij}(k) = \tau,\\
		0, & \text{otherwise}.
	\end{array}\right.
\end{align}
Then, one can rewrite the delay-tolerant version of the LG-type dynamics to solve resource allocation  as,
\begin{align}\label{eq_sol_delay}
	x_i (k+1&) = x_i(k) -\alpha \sum_{j \in \mc{N}^-_i} \sum_{r=0}^{\overline{\tau}} W_{ij} (\partial_{x_i}f_i(k-r) - \partial_{x_j}f_j(k-r)) \mc{I}_{k-r,ij}(r).
\end{align}
Note that to avoid any feasibility gap, the weighted difference of the gradients over every link $(j,i)$  is considered. For all-time feasibility some more assumptions on the delay are needed: the network should be undirected and $\tau_{ij}(k)=\tau_{ji}(k)$. The delays at different links could be arbitrary, heterogeneous, and time-varying in general. Convergence over a nonlinear setup can be established for the following delay-tolerant dynamics (as demonstrated in \cite{ojcsys})
\begin{align} \nonumber
	x_i (k+1) = &x_i(k) \\ &-\alpha \sum_{j \in \mc{N}^-_i} \sum_{r=0}^{\overline{\tau}} W_{ij} \Big(g_l(\partial_{x_i}f_i(k-r)) - g_l(\partial_{x_j}f_j(k-r))\Big) \mc{I}_{k-r,ij}(r).
	\label{eq_sol_L_nonlin_delay}
\end{align}

\subsubsection{Networks subject to Packet Drops}
DRA algorithms should be designed to handle large-scale systems with a high number of nodes. However, as the network scales up, maintaining reliable communication between all nodes becomes more challenging. Network reliability issues, such as a high probability of packet loss, can become more prevalent in large-scale distributed systems, potentially impacting the performance and scalability of the optimization and coordination algorithm. In this direction, some works address coordination and convergence in the presence of packet drops in the distributed setups, see for example robust consensus under packet drops \cite{fagnani2009average,vaidya2012robust,liu2015leader,hadjicostis2015robust}.
In DRA setup, convergence under packet drops cannot be easily addressed by the existing ADMM solutions, and therefore, ideal communications are considered, for example, see \cite{banjac2019decentralized,falsone2020tracking}. One reason is the feasibility violation. In the LG-type setup packet drops are addressed based on link removal, see for example \cite{icrom}.

\section{Applications of DRA }
In this section, we provide motivating examples and real-world applications of DRA.
\label{sec_app}
\subsection{Distributed Energy Resource Management over Smart Grids} \label{sec_edp}
Distributed energy resource management refers to the optimization of power generation and allocation across multiple distributed energy resources in an electrical grid. Traditionally, this is a centralized process where a central authority determines the optimal allocation of power generation resources to meet the load demand while minimizing costs \cite{xia2010optimal,chowdhury1990review}. However, in a distributed setting, where there are numerous small-scale generation sources like solar panels, wind turbines, batteries, and microgrids, the traditional centralized economic dispatch suffers from the drawbacks of the centralized solutions and may not be suitable on large-scale.
By enabling individual energy units or small clusters of energy units to autonomously optimize their power generation and consumption decisions, one can decentralise this process. In the decentralized approach, each unit or cluster assesses the local conditions, such as energy prices, local demand, and availability of renewable energy sources, to make optimal decisions regarding their power generation and consumption \cite{elsayed2014fully,firouzbahrami2022finite,cherukuri2014distributed,vtc,pourbabak2017novel,chen2017delay}. By coordinating and optimizing these distributed decisions, the overall system efficiency and cost-effectiveness can be improved.
The solution often relies on advanced control and communication technologies to facilitate information exchange and coordination among the units \cite{derm}. It can incorporate real-time data, predictive models, and optimization algorithms to dynamically adjust power generation and consumption in response to changing conditions.

%Energy Resource Management involves the efficient and effective management of various energy resources within a system or network. This concept extends beyond power generation to encompass a broader range of energy resources, including both supply-side and demand-side resources. ERM aims to optimize the utilization of energy resources, reduce energy waste, improve energy efficiency, and enhance overall system performance.

In the context of distributed systems, this problem focuses on managing a diverse set of resources such as distributed generation (e.g., solar, wind, and biomass), energy storage (e.g., batteries), electric vehicles, demand response programs, and smart grid technologies. The goal is to \textit{coordinate} and optimize the operation of these resources to meet energy demands, improve system reliability, and achieve economic and environmental objectives.
It involves tasks such as forecasting energy generation and consumption, scheduling and dispatching resources, managing energy storage, coordinating charging and discharging, optimizing demand response programs, and integrating renewable energy sources into the grid. Advanced analytics, optimization algorithms, and control systems are often utilized to enable efficient decision-making and resource coordination.
Other factors like grid constraints, environmental considerations, market dynamics, and regulatory policies ensure the optimal utilization of energy resources while maintaining grid stability and meeting sustainability goals \cite{boqiang2009review}.

\textbf{Mathematical Model:}
The distributed optimization problem for EDP is formulated as minimizing the sum of local energy cost functions \cite{derm,scl} (subject to the equality-constraint mentioned later):
\begin{align} \label{eq_derm}
	\min_{\mb{x},\mb{y}}
	~~ & F(\mb{x},\mb{y}) = \sum_{i=1}^{m} f_i(x_i) + \sum_{j=m+1}^{n} e_j(y_j),
\end{align}
with  $x_i,y_j \in \mathbb{R}$ as the power state of the generator node $i \in \{1,\dots,m\}$ and battery node $j \in \{m+1,\dots,n\}$. Following the formulation~\eqref{eq_dra0} vector $\mb{z} = [x_1;\dots;x_{m};y_{m+1};\dots;y_{n}]$ denotes the global state,  $\widetilde{f}_i(z_i) = f_i(x_i)$ as the local cost of generators and $\widetilde{f}_j(z_j) = e_j(y_j)$ as the local cost of battery nodes. The feasibility constraint coupling the energy resources and the demand is the following equality-constraint:
\begin{align} \label{eq_b}
	\sum_{i=1}^m x_i  = b_{dem} + \sum_{j=m+1}^n y_j.
\end{align}	
This implies that the summation of the produced powers equals the demand  $b_{dem} \in \mathbb{R}$ plus the summation of the reserved powers. This \textit{energy resource-demand feasibility} constraint is of crucial importance and needs to hold at all times; otherwise, the generated and reserved power balance is violated which may cause service disruption and even system break-down. Further, the local box constraints on the energy nodes imply the limiting range of produced/reserved powers at generators/batteries,	
\begin{align} \label{eq_box}
	\underline{x}_i \leq x_i \leq \overline{x}_i, ~\underline{y}_j \leq y_j \leq \overline{y}_j.
\end{align}
It should be recalled from KKT condition in Section~\ref{sec_pre}, from the formulation~\eqref{eq_dra0} and the optimal state \eqref{eq_optimalZ}, one can replace $\mb{a}=[\mb{1}_m,-\mb{1}_{n-m}]$ in the optimal gradient formulation in \eqref{eq_optimalZ}.

\subsection{Distributed CPU Scheduling and Task Allocation over Data Centers}
Distributed CPU scheduling and task allocation over data centres involve the efficient allocation of computational tasks across multiple CPUs or computing nodes/servers in a distributed computing environment.
In such a system, there are multiple CPUs or computing nodes that can execute computational tasks \cite{rikos2022distributed,grammenos2023cpu}. Distributed CPU scheduling aims to optimize the allocation of tasks to these CPUs in order to achieve efficient resource utilization, load balancing, and overall system performance.
Traditional centralized CPU scheduling algorithms, such as those used in standalone operating systems, are not directly applicable in distributed environments due to the distributed nature of resources and the need for inter-node communication. In distributed CPU scheduling, the task allocation decisions are typically made in a decentralized manner, where individual computing nodes collaborate to make scheduling decisions based on local information.

Distributed CPU scheduling algorithms often consider factors such as CPU load, resource availability, network latency, and task characteristics (e.g., priority, estimated execution time) to determine the most suitable node for task execution. The goal is to evenly distribute the workload among the available CPUs, minimize task response time, and optimize resource utilization \cite{kalyvianaki2009self,makridis2020robust,vlahakis2021aimd,chen2022resource}.
On the other hand, task allocation over data centres further refers to the assignment of computational tasks to specific data centres or cloud resources
(see an example in \cite{hattab2019optimized}).
In a large-scale distributed computing environment, data centres are geographically distributed and may have different computing capacities, network connectivity, and service-level agreements. Task allocation aims to optimize the utilization of these resources while satisfying performance objectives and resource constraints \cite{mann2015allocation}.

\textbf{Mathematical Model:}
The data centre is modelled as a set of $n$ computing servers each can also operate as a resource scheduler (which is a standard procedure in modern data centres).
Define the set $\mathcal{J}$ as all tasks to be scheduled and $b_{j} \in \mathcal{J}$ as the task which requires $\rho_{j}$ cycles to be executed.
Additionally, define $T_{h}$ as the time that optimization runs the tasks on the servers.
The CPU capacity during the optimization operation is then equal to $\pi_i^{\max} \coloneqq c_i T_h$, with parameter $c_i$ denoting the sum of all clock rate frequencies of all processing cores at node $i$  \cite{rikos2021optimal}.
%For node $i$, the CPU availability is $\pi_i^{\mathrm{avail}} \coloneqq \pi_i^{\max} - u_{i}$, with $u_{i}$ as the number of unavailable/occupied cycles due to predicted utilization from already running tasks over the period $T_{h}$.
Let the total demanded resources be $\rho \coloneqq \sum_{b_j \in \mathcal{J}} \rho_{j}$.
The cost formulation for this problem is then in standard distributed optimization form \eqref{eq_dra} with the quadratic cost defined as \cite{rikos2021optimal,grammenos2023cpu},
\begin{align} \label{local_cost_w}
	\min_{\mb{x}}
	~~ & F(\mb{x}) = \sum_{i=1}^{n} f_i(x_i) = \sum_{i=1}^{n}\dfrac{1}{2\pi_i^{\max}}  (x_i - \rho_i)^2,
\end{align}
with $x_i$ as the workload to be assigned to node/server $i$ and the workload-demand  constraint (or resource-demand feasibility) as
$\sum_{i=1}^n x_i = \rho$ and some box constraints make the solution non-trivial, in general. These box constraints follow the fact that we need to keep the operating point of the servers away from the overall capacity; in fact, below $70-80\%$ of full capacity (because of the uncertainty of the processing times) since the mean response time of the servers grows (exponentially) at some point \cite{makridis2020robust}. This concern is addressed by box constraints on the load-to-capacity ratios in the form $0 \leq \frac{x_i + \rho_i}{\pi_i^{\max}} \leq 0.75$ as a rule-of-thumb which is modelled via proper extra penalty terms in the objective function \cite{ojcsys}.

\subsection{Distributed Automatic Generation Control} \label{sec_agc}
Distributed automatic generation control is a control mechanism used in power systems to maintain the balance between power generation and load demand across multiple control areas or regions. It involves the coordination and control of various generating units distributed throughout the power grid to regulate the system frequency and ensure stable operation. This problem has some similarities with the economic dispatch problem particularly from an optimization viewpoint and the nature of the optimization variables and cost functions \cite{li2015connecting,boqiang2009review}.
In a power grid, the electricity demand is continuously changing, and the generation must be adjusted in real-time to match the load. Note that, as compared to the economic dispatch problem in Section~\ref{sec_edp}, automatic generation control pertains to the real-time adjustment of the generation outputs of power plants to maintain the balance between supply and demand, ensuring system stability and frequency control. It adjusts the generation outputs based on the deviations between scheduled and actual system conditions. While economic dispatch deals with the \textit{long-term} planning of generation resources, automatic generation control addresses the \textit{short-term} fluctuations and variations in demand and generation, working to maintain the stability and reliability of the power grid in real-time.

Automatic generation control is traditionally implemented in a centralized manner \cite{kumar2005recent}, where a single control centre monitors the system frequency and issues control signals to the generating units to adjust their power output. However, in large interconnected power grids, this centralized control approach may face challenges in terms of communication delays, scalability, and system resilience \cite{kumar2005recent,ullah2021automatic,bevrani2017intelligent}.
Distributed automatic generation control addresses these challenges by decentralizing the control mechanism and allowing individual control areas or regions to autonomously adjust their power generation based on local measurements and information. Each control area has its own processing system, which monitors the local frequency and exchanges information with neighbouring control areas \cite{ullah2021automatic,zhou2020distributed,bevrani2017intelligent}.

\begin{figure}[]
	\centering
	\includegraphics[width=3.5in]{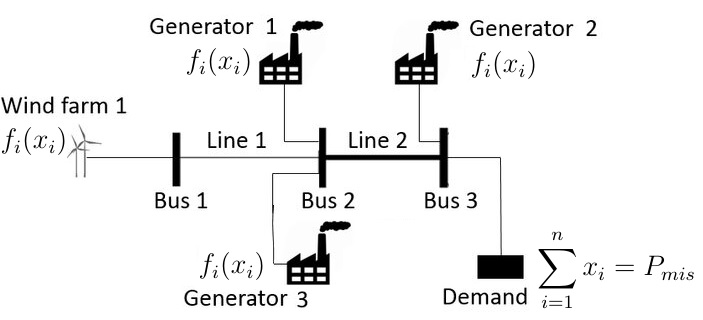}
	\caption{A group of generators producing electricity to meet the load demand: in case there is a power mismatch between the generated power and the demand, the generators readjust their produced power to meet this mismatch while optimizing the sum of local generation cost functions.
	}
	\label{fig_sim2}
\end{figure}

\textbf{Mathematical Model:}
Given the power mismatch, e.g., due to some failed generators, the distributed optimization problem is to cooperatively allocate this power mismatch to the rest of the generators while optimizing the costs. The  problem is as follows:
\begin{align} \label{eq_f_quad}
	\min_\mb{x} &\sum_{i=1}^n f_i(x_i),\\ \nonumber ~\mbox{s.t.}~&\sum_{i=1}^n x_i = P_{mis}, ~~ -\underline{x}_i \leq x_i \leq \overline{x}_i
\end{align}
where $P_{mis}$ is the power mismatch due to the failed generators, $x_i$ is the state of the power state at generator $i$, $\underline{x}_i,\overline{x}_i$  are the upper and lower limits of generated power at the generator $i$. The coupling constraint $\sum_{i=1}^n x_i = P_{mis}$ implies that the mismatch in the power is equal to the new extra allocated powers to the existing generators.
The cost function is typically in the quadratic form $f_i(x_i) = \gamma_i x_i^2+ \beta_i x_i + \alpha_i$ with $\gamma_i,\beta_i,\alpha_i$ defined based on the generator type (fueled by oil, gas, coal, etc.). Some examples of these parameters are given in Table~\ref{tab_par0} \cite{wood2013power,yang2013consensus}.
\begin{table} [hbpt!]
	\centering
	\caption{The cost parameters of the generated powers based on the generator types \cite{wood2013power,yang2013consensus}}
	\begin{tabular}{|c|c|c|c|c|}
		\hline
		Type & $\alpha_i$ (\$/$h$) & $\beta_i$ (\$/$MWh$) & $\gamma_i$ (\$/$MW^2h$) & $\overline{x}_i$ ($MW$)\\
		\hline
		A & 561& 2.0 & 0.04 & 80\\
		\hline
		B& 310& 3.0& 0.03 & 90\\
		\hline
		C &78 & 4.0& 0.035 & 70\\
		\hline
		D& 561& 4.0& 0.03 & 70\\
		\hline
		E &78 & 2.5& 0.04 & 80\\
		\hline		
	\end{tabular}
	\label{tab_par0}
\end{table}

Moreover, in some literature, some realistic ramp-rate-limit (RRL) constraints are also considered, implying that the rate of change of the generated powers is constrained, i.e.,
\begin{align} \label{eq_rrl}
	-R \leq \dot{x}_i \leq R
\end{align}
where $\dot{x}_i$ implies the rate of change in the generator power and $R$ denotes the RRL. This problem follows the standard formulation~\eqref{eq_dra} and, therefore, the optimal condition on the states follows \eqref{eq_optimalX}, i.e., $\mb{a} = \mb{1}_n$ in \eqref{eq_optimalZ}.

Some existing methods in power generation networks in the AGC setup include \cite{cherukuri2015distributed,boyd2006optimal,yang2013consensus,chen2016distributed,yi2016initialization,Kar6345156,zhu2019distributed}. However, these works failed to address the so-called RRLs. More recent solutions address RRLs because of the saturated generator dynamics \cite{doostmohammadian20211st}.

\subsection{Distributed PEV Optimal Charging Schedule}
Distributed PEV (plug-in electric vehicle) optimal charging schedule refers to the decentralized and coordinated scheduling of charging activities for a fleet of electric vehicles. The objective is to optimize the charging schedules of individual vehicles while considering factors such as electricity grid constraints, user preferences, and system-level objectives \cite{sun2017optimal}.  Distributed PEV optimal charging schedule involves individual electric vehicles making charging decisions autonomously based on local information. Instead of a centralized controller dictating the charging schedules for all vehicles, each vehicle independently determines its charging schedule based on its own constraints, preferences, and available information \cite{kisacikoglu2017distributed,mukherjee2016distributed,li2019decentralized}.
The goal of distributed PEV optimal charging schedule is typically to minimize the cost, maximize user convenience, or enhance the overall system efficiency. The optimization objective can vary depending on the specific application and stakeholder requirements. For example, the cost objective may aim to minimize the electricity bill for charging, while the user convenience objective may prioritize charging at preferred times or accommodating specific time constraints \cite{kisacikoglu2017distributed,mukherjee2016distributed,li2019decentralized,abdalrahman2017survey}.
Distributed PEV optimal charging schedule takes into account grid constraints to ensure that charging activities do not overload the electricity grid. By considering factors such as peak demand periods, available grid capacity, and grid stability, the charging schedules can be adjusted to prevent excessive load on the grid and promote grid reliability. Additionally, distributed charging schedules can also incorporate demand response mechanisms, allowing vehicles to respond to dynamic pricing signals or grid signals for load management purposes \cite{dallinger2012grid,amin2020review}.
Various optimization algorithms can be employed by electric vehicles to determine their charging schedules. These algorithms consider the available information, constraints, and objectives to find an optimal or near-optimal charging schedule \cite{esmaili2014optimal,falsone2017dual}.
%Common optimization techniques include linear programming, dynamic programming, stochastic optimization, and game theory.

\textbf{Mathematical Model:}
A fleet of $n$ electric vehicles needs to select a schedule to charge their internal batteries via shared charging stations to minimize the electricity costs \cite{falsone2020tracking,falsone2017dual,vujanic2016decomposition,deori2016decentralized}. There exist both vehicle-level constraints, e.g., user preferences on the final state of charge and
physical limitations of the batteries, and grid-wide constraints, e.g., maximum deliverable power.
%To create a PEV optimal charging schedule, various factors and data need to be considered, including electricity prices, vehicle battery characteristics, grid load information, and renewable energy availability.
In this section, we omit the details of these constraints (refer to \cite{vujanic2016decomposition} for details) and state the main formulation related to the DRA problem \eqref{eq_dra0}.
Let variable $\mb{y}_i$ include the charging and discharging parameters $u_i,v_i \in \{0,1\}$ and $e_i$ as the charge level of battery $i$. Optimizing whether to charge each vehicle or not to charge, i.e., some states in $\mb{y}_i$ follow $0$-$1$ discrete values implying charging or not-charging of the vehicles, implies a mixed integer programming \cite{vujanic2016decomposition,ioan2021mixed}. Reformulating the problem differently compared to \cite{vujanic2016decomposition}, we aim to optimize the charging rate of the batteries \cite{falsone2020tracking,falsone2017dual,li2020distributed}. Then, the distributed optimization problem is in the form,
\begin{align} \label{eq_fpev}
	\min_{\mb{y}_1,\dots,\mb{y}_n} &\sum_{i=1}^n \gamma_i^\top \mb{y}_i,\\ \nonumber ~\mbox{s.t.}~&\sum_{i=1}^n A_i \mb{y}_i \leq b, ~\mb{y}_i \in \mc{Y}_i
\end{align}
with $\mc{Y}_i$ as a set determining the local constraint at node $i$.

Note that the coupling constraint in \eqref{eq_fpev} is an inequality constraint. To reformulate the problem in the standard DRA formulation, consider auxiliary or slack variable $s_i$ in the coupling constraint \cite{falsone2020tracking}. Then, the problem~\eqref{eq_fpev} changes to the standard form as
\begin{align} \label{eq_fpev2}
	\min_{\mb{y}_1,\dots,\mb{y}_n} &\sum_{i=1}^n \gamma_i^\top \mb{y}_i,\\ \nonumber ~\mbox{s.t.}~&\sum_{i=1}^n A_i \mb{y}_i + s_i= b, ~\mb{y}_i \in \mc{Y}_i, ~ 0\leq s_i \leq \overline{s}
\end{align}
Then, the identification variable is $\mb{x}_i = [\mb{y}_i;s_i]$ and the cost function is $f_i(\mb{x}_i) = \gamma_i^\top \mb{y}_i$.

\subsection{Distributed Network Utility Maximization}
Network utility maximization (NUM) is a fundamental concept in the field of network optimization, focusing on the efficient allocation of resources and the maximization of system-wide utility in communication networks \cite{chiang2007layering}. In this framework, the resources, such as bandwidth, power, or capacity, are distributed among network users or components to maximize overall utility or performance.
Each user or component in the network is associated with a utility function that quantifies their preference for allocated resources.
This objective is to maximize the aggregate utility of all users or components in the network while addressing resource constraints such as bandwidth limits, energy constraints, or capacity restrictions.
These constraints need to be satisfied while optimizing the objective.
This objective is to maximize the aggregate utility of all users or components in the network while addressing the resource constraints.

NUM problems are often complex and high-dimensional, making direct optimization challenging. Decomposition methods are used to break down the problem into smaller, more manageable sub-problems, allowing for efficient optimization \cite{palomar2006tutorial}.
NUM plays a critical role in various applications including, telecommunication networks \cite{lin2006tutorial,srikant2013communication}, spectrum allocation \cite{hasan2016network,tsiropoulos2014radio}, network energy management \cite{kraning2013dynamic}, Internet content delivery \cite{trossen2012designing,ercetin2003market}, and Cloud computing \cite{guo2018energy,ergu2013analytic}. A review of more recent applications in Fog and IoT networks is provided by \cite{chiang2016fog}.

\textbf{Mathematical Model:}
This problem models a communication network with a set of $\mc{L}$ links, where the link/channel has certain finite capacity denoted by $c$. These channels are shared with a set of $n$ sources or nodes denoted by $s$. Each source in $s$ uses a set of $\mc{L}(s) \in \mc{L}$ of links with certain utility function $U_s(x_s)$ where state $x_s$ denotes the transmission rate of the source.
%In fact, the set $\mc{L}(s)$ denotes a routing matrix.
The network utility maximization is modelled as
\cite{chiang2007layering,kelly1998rate}
\begin{align} \label{eq_num}	\max_{\mb{x}_1,\dots,\mb{x}_n} &\sum_s U_s \mb{x}_s,\\ \nonumber ~\mbox{s.t.}~&R \mb{x} \leq c
\end{align}
with $R$ as network congestion protocol defined as:
\begin{align}
	R_{ls} = \left\{ \begin{array}{ll}
		1, & \text{if $l \in \mc{L}(s)$ source $s$ uses link $l$},\\
		0, & \text{otherwise}.
	\end{array}\right.
\end{align}
Similar to \eqref{eq_fpev} this problem can be reformulated to standard DRA formulation as in \eqref{eq_fpev2}, by use of auxiliary/slack variables in the coupling constraint.

\subsection{Other Applications}
There are many other diverse applications of DRA where the formulation is a bit different compared to the above-mentioned cases. Some examples of these applications are listed next where efficient resource allocation and coordination among distributed entities are essential for optimal system performance and resource utilization.

\subsubsection{Multi-Robot Coordination Systems for Coverage}
Distributed resource allocation is crucial in multi-robot systems where multiple robots collaborate to perform tasks in complex environments. It involves allocating tasks and resources, such as sensing capabilities, processing power, and sensor coverage, to different robots to optimize the overall cost/objective, achieve task completion, and ensure coordination among the robots \cite{msc09,MiadCons,cortes2004coverage,bullo2009distributed,telsang2022decentralized}.

\subsubsection{Transportation and Logistics}
In transportation and logistics systems, DRA is essential for optimizing the allocation of vehicles, routes, and delivery tasks. It involves distributing tasks to different vehicles based on their locations, capacities, and availability to minimize delivery time, reduce fuel consumption, and improve overall efficiency \cite{baldacci2012recent,taillard1996vehicle,gao2021priority}.

%\subsubsection{Wireless Sensor Networks:} In wireless sensor networks, DRA is crucial for the efficient utilization of limited resources such as energy, bandwidth, and computational capabilities. It involves tasks such as distributed data aggregation, routing, power control, and channel allocation to optimize resource utilization and prolong network lifetime \cite{heinzelman2000energy,akyildiz2002survey}.

\subsubsection{Spreading Processes and Control of Epidemic Disease}
In the spread of viruses or diseases over a network, different models for epidemic spreading are considered, where the most common model is linearized Susceptible-Infected-Susceptible (SIS). A distributed application over multi-agent social networks is to locally compute the optimal allocation of vaccination resources (for prevention) and antidotes (for treatment) to control the contagion
\cite{enyioha2015distributed,nowzari2016analysis}.

\section{Conclusion and Future Directions} \label{sec_conclusion}
This paper presents a comprehensive review of distributed algorithms for resource allocation over multi-agent systems, including mathematical formulations, relevant applications, various existing solutions, and algorithmic properties. The exploration of these topics has shed light on the challenges and potential solutions in this domain, providing a better comparative analysis of the existing strategies. Specifically, the pros and cons of the primal and dual optimization methods are discussed in detail, giving insight into when/how to apply each method. The comparative analysis in terms of feasibility, convergence, and network properties further enlightens the specific applications and also limitations of each method. Our thorough analysis of the algorithmic properties enables readers to understand the trade-offs involved in employing different distributed algorithms. By comparing convergence rates, communication networking overhead, and constraint feasibility, we provide nuanced insights into the strengths and weaknesses of each approach.

As the field of distributed resource allocation continues to progress, numerous avenues for future research emerge, opening new opportunities for advancements in this domain. Investigating dynamic adaptation in algorithms to accommodate changing system conditions, such as fluctuating resource availability, (mobile) agent arrivals/departures, or evolving demand patterns, presents a promising focus. Areas like asynchronous scheduling \cite{themis2021cpu}, robust and noise-resilient solutions \cite{wei2017consensus,stankovic2019robust}, and low-bit communication-efficient strategies \cite{taes2020finite} stand as examples for future exploration. Developing adaptive resource allocation strategies holds the potential to create more responsive and efficient systems that effectively handle dynamic environments. Furthermore, integrating machine learning and artificial intelligence techniques into resource allocation algorithms could enhance decision-making processes significantly. Exploring reinforcement learning, deep learning, or evolutionary algorithms \cite{peteiro2013survey,kumar2022machine}. offers an exciting prospect to improve resource allocation efficiency and adaptability. Additionally, the pursuit of data-based modelling and optimization \cite{mohajer2017big} holds promise as an interesting research direction. Finally, given the rising emphasis on energy efficiency and sustainability, future research could focus on resource allocation algorithms explicitly incorporating energy consumption (green computing) and environmental impact as optimization criteria.

\bibliographystyle{elsarticle-num}
\bibliography{bibliography}

\begin{thebibliography}{100}
\expandafter\ifx\csname url\endcsname\relax
  \def\url#1{\texttt{#1}}\fi
\expandafter\ifx\csname urlprefix\endcsname\relax\def\urlprefix{URL }\fi
\expandafter\ifx\csname href\endcsname\relax
  \def\href#1#2{#2} \def\path#1{#1}\fi

\bibitem{dibaji2019systems}
S.~M. Dibaji, M.~Pirani, D.~B. Flamholz, A.~M. Annaswamy, K.~H. Johansson,
  A.~Chakrabortty, A systems and control perspective of {CPS} security, Annual
  reviews in control 47 (2019) 394--411.

\bibitem{jstsp}
M.~Doostmohammadian, U.~Khan, On the genericity properties in distributed
  estimation: Topology design and sensor placement, IEEE Journal of Selected
  Topics in Signal Processing 7~(2) (2013) 195--204.

\bibitem{kar2013consensus}
S.~Kar, J.~M.~F. Moura, Consensus+ innovations distributed inference over
  networks: cooperation and sensing in networked systems, IEEE Signal
  Processing Magazine 30~(3) (2013) 99--109.

\bibitem{nuno-suff.ness}
S.~Park, N.~Martins, Necessary and sufficient conditions for the
  stabilizability of a class of {LTI} distributed observers, in: 51st IEEE
  Conference on Decision and Control, 2012, pp. 7431--7436.

\bibitem{jadbabaie2003coordination}
A.~Jadbabaie, J.~Lin, A.~S. Morse, Coordination of groups of mobile autonomous
  agents using nearest neighbor rules, IEEE Transactions on automatic control
  48~(6) (2003) 988--1001.

\bibitem{cps_bullo}
F.~Pasqualetti, F.~Dorfler, F.~Bullo, Attack detection and identification in
  cyber-physical systems, IEEE Transactions on Automatic Control 58~(11) (2013)
  2715--2729.
\newblock \href {https://doi.org/10.1109/TAC.2013.2266831}
  {\path{doi:10.1109/TAC.2013.2266831}}.

\bibitem{khan2020optimization}
U.~A. Khan, W.~U. Bajwa, A.~Nedi{\'c}, M.~G. Rabbat, A.~H. Sayed, Optimization
  for data-driven learning and control, Proceedings of the IEEE 108~(11) (2020)
  1863--1868.

\bibitem{molzahn2017survey}
D.~K. Molzahn, F.~D{\"o}rfler, H.~Sandberg, S.~H. Low, S.~Chakrabarti,
  R.~Baldick, J.~Lavaei, A survey of distributed optimization and control
  algorithms for electric power systems, IEEE Transactions on Smart Grid 8~(6)
  (2017) 2941--2962.

\bibitem{yang2013consensus}
S.~Yang, S.~Tan, J.~Xu, Consensus based approach for economic dispatch problem
  in a smart grid, IEEE Transactions on Power Systems 28~(4) (2013) 4416--4426.

\bibitem{themis2011asynchronous}
C.~N. Hadjicostis, T.~Charalambous, Asynchronous coordination of distributed
  energy resources for the provisioning of ancillary services, in: 49th IEEE
  Allerton Conference on Communication, Control, and Computing, 2011, pp.
  1500--1507.

\bibitem{cherukuri2015distributed}
A.~Cherukuri, J.~Cort{\'e}s, Distributed generator coordination for
  initialization and anytime optimization in economic dispatch, IEEE
  Transactions on Control of Network Systems 2~(3) (2015) 226--237.

\bibitem{pelikan2006scalable}
M.~Pelikan, K.~Sastry, E.~Cant{\'u}-Paz, Scalable optimization via
  probabilistic modeling: From algorithms to applications, Springer, 2006.

\bibitem{vinyals2011survey}
M.~Vinyals, J.~Rodriguez-Aguilar, J.~Cerquides, A survey on sensor networks
  from a multiagent perspective, The Computer Journal 54~(3) (2011) 455--470.

\bibitem{ramseyer2023speedex}
G.~Ramseyer, A.~Goel, D.~Mazi{\`e}res, {SPEEDEX}: A scalable, parallelizable,
  and economically efficient decentralized {EXchange}, in: 20th USENIX
  Symposium on Networked Systems Design and Implementation (NSDI 23), 2023, pp.
  849--875.

\bibitem{skobelev2015multi}
P.~Skobelev, Multi-agent systems for real-time adaptive resource management,
  in: Industrial Agents, Elsevier, 2015, pp. 207--229.

\bibitem{shreyas2020byzantine}
K.~Kuwaranancharoen, L.~Xin, S.~Sundaram, Byzantine-resilient distributed
  optimization of multi-dimensional functions, in: American Control Conference,
  IEEE, 2020, pp. 4399--4404.

\bibitem{pirani2018cooperative}
M.~Pirani, E.~Hashemi, A.~Khajepour, B.~Fidan, B.~Litkouhi, S.~Chen,
  S.~Sundaram, Cooperative vehicle speed fault diagnosis and correction, IEEE
  Transactions on Intelligent Transportation Systems 20~(2) (2018) 783--789.

\bibitem{pequito2014design}
S.~Pequito, S.~Kar, S.~Sundaram, A.~P. Aguiar, Design of communication networks
  for distributed computation with privacy guarantees, in: IEEE 53rd Annual
  Conference on Decision and Control (CDC), 2014, pp. 1370--1376.

\bibitem{rikos_quant}
A.~I. Rikos, T.~Charalambous, K.~H. Johansson, C.~N. Hadjicostis,
  Privacy-preserving event-triggered quantized average consensus, in: 59th IEEE
  Conference on Decision and Control, 2020, pp. 6246--6253.

\bibitem{ruan2019secure}
M.~Ruan, H.~Gao, Y.~Wang, Secure and privacy-preserving consensus, IEEE
  Transactions on Automatic Control 64~(10) (2019) 4035--4049.

\bibitem{mo2016privacy}
Y.~Mo, R.~M. Murray, Privacy preserving average consensus, IEEE Transactions on
  Automatic Control 62~(2) (2016) 753--765.

\bibitem{nekouei2019information}
E.~Nekouei, T.~Tanaka, M.~Skoglund, K.~H. Johansson, Information-theoretic
  approaches to privacy in estimation and control, Annual Reviews in Control 47
  (2019) 412--422.

\bibitem{dsvm}
M.~Doostmohammadian, A.~Aghasi, T.~Charalambous, U.~A. Khan, Distributed
  support vector machine over dynamic balanced directed networks, IEEE Control
  Systems Letters 6 (2021) 758--763.

\bibitem{heinzelman2000energy}
W.~R. Heinzelman, A.~Chandrakasan, H.~Balakrishnan, Energy-efficient
  communication protocol for wireless microsensor networks, in: Proceedings of
  the 33rd Annual Hawaii International Conference on System Sciences, IEEE,
  2000, pp. 10--pp.

\bibitem{akyildiz2002survey}
I.~F. Akyildiz, W.~Su, Y.~Sankarasubramaniam, E.~Cayirci, A survey on sensor
  networks, IEEE Communications magazine 40~(8) (2002) 102--114.

\bibitem{vlahakis2021aimd}
E.~Vlahakis, N.~Athanasopoulos, S.~McLoone, {AIMD} scheduling and resource
  allocation in distributed computing systems, in: 60th IEEE Conference on
  Decision and Control (CDC), IEEE, 2021, pp. 4642--4647.

\bibitem{baldacci2012recent}
R.~Baldacci, A.~Mingozzi, R.~Roberti, Recent exact algorithms for solving the
  vehicle routing problem under capacity and time window constraints, European
  Journal of Operational Research 218~(1) (2012) 1--6.

\bibitem{taillard1996vehicle}
E.~D. Taillard, G.~Laporte, M.~Gendreau, Vehicle routeing with multiple use of
  vehicles, Journal of the Operational research society 47~(8) (1996)
  1065--1070.

\bibitem{gao2021priority}
R.~Gao, H.~Niu, A priority-based {ADMM} approach for flexible train scheduling
  problems, Transportation Research Part C: Emerging Technologies 123 (2021)
  102960.

\bibitem{boqiang2009review}
R.~Boqiang, J.~Chuanwen, A review on the economic dispatch and risk management
  considering wind power in the power market, Renewable and Sustainable Energy
  Reviews 13~(8) (2009) 2169--2174.

\bibitem{MiadCons}
H.~Sayyaadi, M.~Moarref, A distributed algorithm for proportional task
  allocation in networks of mobile agents, IEEE Transactions on Automatic
  Control 56~(2) (2011) 405--410.

\bibitem{zander1997radio}
J.~Zander, Radio resource management in future wireless networks: Requirements
  and limitations, IEEE Communications magazine 35~(8) (1997) 30--36.

\bibitem{lesser2003distributed}
V.~Lesser, C.~L. Ortiz~Jr, M.~Tambe, Distributed sensor networks: A multiagent
  perspective, Vol.~9, Springer Science \& Business Media, 2003.

\bibitem{malairajan2013class}
R.~A. Malairajan, K.~Ganesh, M.~Muhos, S.~P. Anbuudayasankar, Class of resource
  allocation problems in supply chain--a review, International Journal of
  Business Innovation and Research 7~(1) (2013) 113--139.

\bibitem{nedic2018distributed}
A.~Nedi{\'c}, J.~Liu, Distributed optimization for control, Annual Review of
  Control, Robotics, and Autonomous Systems 1 (2018) 77--103.

\bibitem{yang2019distributed}
P.~Yang, M.~Xu, D.~Li, Z.~Liu, Y.~Huang, Distributed fault tolerant consensus
  control for multi-agent system with actuator fault based on adaptive
  observer, Transactions of the Institute of Measurement and Control (2019)
  1--11.

\bibitem{doostmohammadian20211st}
M.~Doostmohammadian, A.~Aghasi, M.~Vrakopoulou, T.~Charalambous, 1st-order
  dynamics on nonlinear agents for resource allocation over uniformly-connected
  networks, in: IEEE Conference on Control Technology and Applications, 2022,
  pp. 1184--1189.

\bibitem{lakshmanan2008decentralized}
H.~Lakshmanan, D.~P. De~Farias, Decentralized resource allocation in dynamic
  networks of agents, SIAM Journal on Optimization 19~(2) (2008) 911--940.

\bibitem{bertsekas1975necessary}
D.~P. Bertsekas, Necessary and sufficient conditions for a penalty method to be
  exact, Mathematical programming 9~(1) (1975) 87--99.

\bibitem{nesterov1998introductory}
Y.~Nesterov, Introductory lectures on convex programming, {volume I}: Basic
  course, Lecture notes 3~(4) (1998) 5.

\bibitem{mikael2021cdc}
X.~Wu, S.~Magnusson, M.~Johansson, A new family of feasible methods for
  distributed resource allocation, in: 60th IEEE Conference on Decision and
  Control, 2021, pp. 3355--3360.

\bibitem{cortes2008discontinuous}
J.~Cortes, Discontinuous dynamical systems, IEEE Control systems magazine
  28~(3) (2008) 36--73.

\bibitem{bertsekas_lecture}
D.~P. Bertsekas, A.~Nedic, A.~E. Ozdaglar, Convexity, duality, and lagrange
  multipliers, Lecture Notes, MIT Press (2001).

\bibitem{boyd2006optimal}
L.~Xiao, S.~Boyd, Optimal scaling of a gradient method for distributed resource
  allocation, Journal of Optimization Theory and Applications 129~(3) (2006)
  469--488.

\bibitem{lj_lecture}
A.~Iouditski, Convex optimization {I}: Introduction, Lecture Notes, Laboratoire
  Jean Kuntzmann, FR (2015).

\bibitem{yang2019survey}
T.~Yang, X.~Yi, J.~Wu, Y.~Yuan, D.~Wu, Z.~Meng, Y.~Hong, H.~Wang, Z.~Lin, K.~H.
  Johansson, A survey of distributed optimization, Annual Reviews in Control 47
  (2019) 278--305.

\bibitem{godsil}
C.~Godsil, G.~Royle, Algebraic graph theory, New York: Springer, 2001.

\bibitem{olfati_rev}
R.~Olfati-Saber, J.~A. Fax, R.~M. Murray, Consensus and cooperation in
  networked multi-agent systems, Proceedings of the IEEE 95~(1) (2007)
  215--233.

\bibitem{zhang2020distributed}
J.~Zhang, K.~You, K.~Cai, Distributed dual gradient tracking for resource
  allocation in unbalanced networks, IEEE Transactions on Signal Processing 68
  (2020) 2186--2198.

\bibitem{gharesifard2012distributed}
B.~Gharesifard, J.~Cort{\'e}s, Distributed strategies for generating
  weight-balanced and doubly stochastic digraphs, European Journal of Control
  18~(6) (2012) 539--557.

\bibitem{hadjicostis2018distributed}
C.~N. Hadjicostis, A.~Dom{\'\i}nguez-Garc{\'\i}a, T.~Charalambous, et~al.,
  Distributed averaging and balancing in network systems: with applications to
  coordination and control, Foundations and Trends{\textregistered} in Systems
  and Control 5~(2-3) (2018) 99--292.

\bibitem{nedic2014distributed}
A.~Nedi{\'c}, A.~Olshevsky, Distributed optimization over time-varying directed
  graphs, IEEE Transactions on Automatic Control 60~(3) (2014) 601--615.

\bibitem{ren2005consensus}
W.~Ren, R.~W. Beard, Consensus seeking in multiagent systems under dynamically
  changing interaction topologies, IEEE Transactions on Automatic Control
  50~(5) (2005) 655--661.

\bibitem{falsone2020tracking}
A.~Falsone, I.~Notarnicola, G.~Notarstefano, M.~Prandini, Tracking-{ADMM} for
  distributed constraint-coupled optimization, Automatica 117 (2020) 108962.

\bibitem{rikos2021optimal}
A.~I. Rikos, A.~Grammenos, E.~Kalyvianaki, C.~N. Hadjicostis, T.~Charalambous,
  K.~H. Johansson, Optimal {CPU} scheduling in data centers via a finite-time
  distributed quantized coordination mechanism, in: 60th IEEE Conference on
  Decision and Control, 2021, pp. 6276--6281.

\bibitem{banjac2019decentralized}
G.~Banjac, F.~Rey, P.~Goulart, J.~Lygeros, Decentralized resource allocation
  via dual consensus {ADMM}, in: IEEE American Control Conference, 2019, pp.
  2789--2794.

\bibitem{6426252}
N.~H. Vaidya, C.~N. Hadjicostis, A.~D. Dominguez-Garcia, Robust average
  consensus over packet dropping links: Analysis via coefficients of
  ergodicity, in: 51st IEEE Conference on Decision and Control, 2012, pp.
  2761--2766.

\bibitem{SensNets:Olfati04}
R.~Olfati-Saber, R.~M. Murray, Consensus problems in networks of agents with
  switching topology and time-delays, IEEE Transactions on Automatic Control
  49, no. 9 (2004) 1520--1533.

\bibitem{graph_handbook}
J.~L. Gross, J.~Yellen, Handbook of Graph Theory, CRC Press, 2004.

\bibitem{wang2008algebraic}
H.~Wang, P.~Van~Mieghem, Algebraic connectivity optimization via link addition,
  in: 3rd International Conference on Bio-Inspired Models of Network,
  Information and Computing Systems, 2008, pp. 1--8.

\bibitem{ojcsys}
M.~Doostmohammadian, A.~Aghasi, A.~I. Rikos, A.~Grammenos, E.~Kalyvianaki,
  C.~N. Hadjicostis, K.~H. Johansson, T.~Charalambous, Distributed
  anytime-feasible resource allocation subject to heterogeneous time-varying
  delays, IEEE Open Journal of Control Systems 1 (2022) 255--267.

\bibitem{renetal05}
W.~Ren, R.~W. Beard, E.~M. Atkins, A survey of consensus problems in
  multi-agent coordination, in: American Control Conference, Portland, OR,
  2005, pp. 1859--1864.

\bibitem{cao2015leader}
W.~Cao, J.~Zhang, W.~Ren, Leader--follower consensus of linear multi-agent
  systems with unknown external disturbances, Systems \& Control Letters 82
  (2015) 64--70.

\bibitem{tanner02}
H.~G. Tanner, G.~J. Pappas, V.~Kumar, Leader-to-formation stability, IEEE
  Transactions on Robotics and Automation 20~(3) (2004) 433--455.

\bibitem{xiao2008asynchronous}
F.~Xiao, L.~Wang, Asynchronous consensus in continuous-time multi-agent systems
  with switching topology and time-varying delays, IEEE Transactions on
  Automatic Control 53~(8) (2008) 1804--1816.

\bibitem{kia2019tutorial}
S.~S. Kia, B.~Van~Scoy, J.~Cortes, R.~A. Freeman, K.~M. Lynch, S.~Martinez,
  Tutorial on dynamic average consensus: The problem, its applications, and the
  algorithms, IEEE Control Systems Magazine 39~(3) (2019) 40--72.

\bibitem{msc09}
M.~Doostmohammadian, H.~Sayyaadi, M.~Moarref, A novel consensus protocol using
  facility location algorithms, in: IEEE Conference on Control Applications \&
  Intelligent Control, 2009, pp. 914--919.

\bibitem{cortes2004coverage}
J.~Cort{\'e}s, S.~Mart{\i}nez, T.~Karatas, F.~Bullo, Coverage control for
  mobile sensing networks, IEEE Transactions on Robotics and Automation 20~(2)
  (2004).

\bibitem{bullo2009distributed}
F.~Bullo, J.~Cort{\'e}s, S.~Martinez, Distributed control of robotic networks:
  a mathematical approach to motion coordination algorithms, Vol.~27, Princeton
  University Press, 2009.

\bibitem{telsang2022decentralized}
B.~Telsang, Decentralized resource allocation through constrained centroidal
  voronoi tessellations, Ph.{D}. thesis, The University of Tennessee --
  Knoxville (2022).

\bibitem{bullo-opinion}
A.~Mirtabatabaei, F.~Bullo, Opinion dynamics in heterogeneous networks:
  Convergence conjectures and theorems, SIAM Journal on Control and
  Optimization 50~(5) (2012) 2763--2785.

\bibitem{jstsp14}
M.~Doostmohammadian, U.~Khan, Graph-theoretic distributed inference in social
  networks, IEEE Journal of Selected Topics in Signal Processing 8~(4) (2014)
  613--623.

\bibitem{cortes2006robust}
J.~Cort{\'e}s, S.~Mart{\'\i}nez, F.~Bullo, Robust rendezvous for mobile
  autonomous agents via proximity graphs in arbitrary dimensions, IEEE
  Transactions on Automatic Control 51~(8) (2006) 1289--1298.

\bibitem{Jadbabaie_flocking}
H.~Tanner, A.~Jadbabaie, G.~J. Pappas, Flocking in fixed and switching
  networks, IEEE Transactions on Automatic Control 52~(5) (2007) 863--868.

\bibitem{olfati2002distributed}
R.~Olfati-Saber, R.~M. Murray, Distributed cooperative control of multiple
  vehicle formations using structural potential functions, IFAC Proceedings
  Volumes 35~(1) (2002) 495--500.

\bibitem{flock}
R.~Olfati-Saber, P.~Jalalkamali, Collaborative target tracking using
  distributed kalman filtering on mobile sensor networks, in: American Control
  Conference, San Francisco, CA, 2011.

\bibitem{taes}
M.~Doostmohammadian, A.~Taghieh, H.~Zarrabi, Distributed estimation approach
  for tracking a mobile target via formation of {UAV}s, IEEE Transactions on
  Automation Science and Engineering 19~(4) (2021) 3765--3776.

\bibitem{tcns_fdi}
M.~Doostmohammadian, N.~Meskin, Sensor fault detection and isolation via
  networked estimation: Full-rank dynamical systems, IEEE Transactions on
  Control of Network Systems 8~(2) (2020) 987--996.

\bibitem{ijc_fdi}
M.~Doostmohammadian, H.~Zarrabi, T.~Charalambous, Sensor fault detection and
  isolation via networked estimation: rank-deficient dynamical systems,
  International Journal of Control (2022) 1--18.

\bibitem{giraldo2018survey}
J.~Giraldo, D.~Urbina, A.~Cardenas, J.~Valente, M.~Faisal, J.~Ruths,
  N.~Tippenhauer, H.~Sandberg, R.~Candell, A survey of physics-based attack
  detection in cyber-physical systems, ACM Computing Surveys (CSUR) 51~(4)
  (2018) 1--36.

\bibitem{deghat2019detection}
M.~Deghat, V.~Ugrinovskii, I.~Shames, C.~Langbort, Detection and mitigation of
  biasing attacks on distributed estimation networks, Automatica 99 (2019)
  369--381.

\bibitem{xin2020decentralized}
R.~Xin, S.~Kar, U.~A. Khan, Decentralized stochastic optimization and machine
  learning: A unified variance-reduction framework for robust performance and
  fast convergence, IEEE Signal Processing Magazine 37~(3) (2020) 102--113.

\bibitem{sherayas:08}
S.~Sundaram, C.~Hadjicostis, Distributed function calculation and consensus
  using linear iterative strategies, IEEE Journal on Selected Areas in
  Communications 26~(4) (2008) 650--660.

\bibitem{taes2020finite}
M.~Doostmohammadian, Single-bit consensus with finite-time convergence: Theory
  and applications, IEEE Transactions on Aerospace and Electronic Systems
  56~(4) (2020) 3332--3338.

\bibitem{feng2017finite}
Z.~Feng, G.~Hu, Finite-time distributed optimization with quadratic objective
  functions under uncertain information, in: IEEE 56th Annual Conference on
  Decision and Control (CDC), IEEE, 2017, pp. 208--213.

\bibitem{hu2018distributed}
Z.~Hu, J.~Yang, Distributed finite-time optimization for second order
  continuous-time multiple agents systems with time-varying cost function,
  Neurocomputing 287 (2018) 173--184.

\bibitem{lin2016distributed}
P.~Lin, W.~Ren, J.~A. Farrell, Distributed continuous-time optimization:
  nonuniform gradient gains, finite-time convergence, and convex constraint
  set, IEEE Transactions on Automatic Control 62~(5) (2017) 2239--2253.

\bibitem{bhat2000finite}
S.~P. Bhat, D.~S. Bernstein, Finite-time stability of continuous autonomous
  systems, SIAM Journal on Control and Optimization 38~(3) (2000) 751--766.

\bibitem{Scientia2011}
H.~Sayyaadi, M.~Doostmohammadian, Finite-time consensus in directed switching
  network topologies and time-delayed communications, Scientia Iranica 18~(1)
  (2011) 75--85.

\bibitem{Garg_fixed}
K.~Garg, D.~Panagou, Fixed-time stable gradient flows: Applications to
  continuous-time optimization, IEEE Transactions on Automatic Control 66~(5)
  (2021) 2002--2015.

\bibitem{shang2017}
Y.~Shang, Fixed-time group consensus for multi-agent systems with non-linear
  dynamics and uncertainties, IET Control Theory \& Applications (2017).

\bibitem{ning2017distributed}
B.~Ning, Q.~Han, Z.~Zuo, Distributed optimization for multiagent systems: An
  edge-based fixed-time consensus approach, IEEE Transactions on Cybernetics
  49~(1) (2017) 122--132.

\bibitem{firouzbahrami2022cooperative}
M.~Firouzbahrami, A.~Nobakhti, Cooperative fixed-time/finite-time distributed
  robust optimization of multi-agent systems, Automatica 142 (2022) 110358.

\bibitem{wang2018prescribed}
Y.~Wang, Y.~Song, D.~J. Hill, M.~Krstic, Prescribed-time consensus and
  containment control of networked multiagent systems, IEEE transactions on
  cybernetics 49~(4) (2018) 1138--1147.

\bibitem{ji2023initialization}
L.~Ji, L.~Yu, C.~Zhang, X.~Guo, H.~Li, Initialization-free distributed
  prescribed-time consensus based algorithm for economic dispatch problem over
  directed network, Neurocomputing 533 (2023) 1--9.

\bibitem{kia2015distributed}
S.~S. Kia, J.~Cort{\'e}s, S.~Mart{\'\i}nez, Distributed convex optimization via
  continuous-time coordination algorithms with discrete-time communication,
  Automatica 55 (2015) 254--264.

\bibitem{Uribe2020}
C.~A. Uribe, S.~Lee, A.~Gasnikov, A.~Nedic, A dual approach for optimal
  algorithms in distributed optimization over networks, in: Information Theory
  and Applications Workshop, 2020, pp. 1--37.
\newblock \href {https://doi.org/10.1109/ITA50056.2020.9244951}
  {\path{doi:10.1109/ITA50056.2020.9244951}}.

\bibitem{xin2019frost}
R.~Xin, C.~Xi, U.~A. Khan, {FROST}--fast row-stochastic optimization with
  uncoordinated step-sizes, EURASIP Journal on Advances in Signal Processing
  2019~(1) (2019) 1--14.

\bibitem{qureshi2020s}
M.~I. Qureshi, R.~Xin, S.~Kar, U.~A. Khan, {S-ADDOPT}: Decentralized stochastic
  first-order optimization over directed graphs, IEEE Control Systems Letters
  5~(3) (2020) 953--958.

\bibitem{agarwal2012delay}
A.~Agarwal, J.~C. Duchi, Distributed delayed stochastic optimization, in: 51st
  IEEE Conference on Decision and Control, 2012, pp. 5451--5452.

\bibitem{al2020gradient}
H.~Al-Lawati, S.~C. Draper, Gradient delay analysis in asynchronous distributed
  optimization, in: IEEE International Conference on Acoustics, Speech and
  Signal Processing, 2020, pp. 4207--4211.

\bibitem{gharesifard2013distributed}
B.~Gharesifard, J.~Cort{\'e}s, Distributed continuous-time convex optimization
  on weight-balanced digraphs, IEEE Transactions on Automatic Control 59~(3)
  (2013) 781--786.

\bibitem{shames2011accelerated}
E.~Ghadimi, M.~Johansson, I.~Shames, Accelerated gradient methods for networked
  optimization, in: IEEE American Control Conference, 2011, pp. 1668--1673.

\bibitem{doan2017scl}
T.~T. Doan, A.~Olshevsky, Distributed resource allocation on dynamic networks
  in quadratic time, Systems \& Control Letters 99 (2017) 57--63.

\bibitem{doan2017ccta}
T.~T. Doan, C.~L. Beck, Distributed lagrangian methods for network resource
  allocation, in: IEEE Conference on Control Technology and Applications
  (CCTA), 2017, pp. 650--655.

\bibitem{nedic2018improved}
A.~Nedi{\'c}, A.~Olshevsky, W.~Shi, Improved convergence rates for distributed
  resource allocation, in: IEEE Conference on Decision and Control (CDC), IEEE,
  2018, pp. 172--177.

\bibitem{Turan2021}
B.~Turan, C.~A. Uribe, H.~Wai, M.~Alizadeh, Resilient prima-dual optimization
  algorithms for distributed resource allocation, IEEE Transactions on Control
  of Network Systems 8~(1) (2021) 282--294.
\newblock \href {https://doi.org/10.1109/TCNS.2020.3024485}
  {\path{doi:10.1109/TCNS.2020.3024485}}.

\bibitem{Uribe19}
D.~E. Ochoa, J.~I. Poveda, C.~A. Uribe, N.~Quijano, Hybrid robust optimal
  resource allocation with momentum, in: 58th Conference on Decision and
  Control, 2019, pp. 3954--3959.

\bibitem{sayed2019proximal}
S.~A. Alghunaim, K.~Yuan, A.~H. Sayed, A proximal diffusion strategy for
  multiagent optimization with sparse affine constraints, IEEE Transactions on
  Automatic Control 65~(11) (2019) 4554--4567.

\bibitem{hamedani2017multi}
E.~Yazdandoost~Hamedani, N.~S. Aybat, Multi-agent constrained optimization of a
  strongly convex function, in: Global Conference on Signal and Information
  Processing (GlobalSIP), IEEE, 2017, pp. 558--562.

\bibitem{aybat2016distributed}
N.~S. Aybat, E.~Yazdandoost~Hamedani, Distributed primal-dual method for
  multi-agent sharing problem with conic constraints, in: 50th IEEE Asilomar
  Conference on Signals, Systems and Computers, 2016, pp. 777--782.

\bibitem{yi2016initialization}
P.~Yi, Y.~Hong, F.~Liu, Initialization-free distributed algorithms for optimal
  resource allocation with feasibility constraints and application to economic
  dispatch of power systems, Automatica 74 (2016) 259--269.

\bibitem{wang2018distributed}
D.~Wang, Z.~Wang, M.~Chen, W.~Wang, Distributed optimization for multi-agent
  systems with constraints set and communication time-delay over a directed
  graph, Information Sciences 438 (2018) 1--14.

\bibitem{jiang2021distributed}
W.~Jiang, T.~Charalambous, Distributed alternating direction method of
  multipliers using finite-time exact ratio consensus in digraphs, in: European
  Control Conference (ECC), IEEE, 2021, pp. 2205--2212.

\bibitem{fast}
M.~Doostmohammadian, A.~Aghasi, M.~Pirani, E.~Nekouei, U.~A. Khan,
  T.~Charalambous, Fast-convergent anytime-feasible dynamics for distributed
  allocation of resources over switching sparse networks with quantized
  communication links, in: European Control Conference, 2022, pp. 84--89.

\bibitem{xu2018dual}
J.~Xu, S.~Zhu, Y.~Soh, L.~Xie, A dual splitting approach for distributed
  resource allocation with regularization, IEEE Transactions on Control of
  Network Systems 6~(1) (2018) 403--414.

\bibitem{zhu2019distributed}
Y.~Zhu, W.~Ren, W.~Yu, G.~Wen, Distributed resource allocation over directed
  graphs via continuous-time algorithms, IEEE Transactions on Systems, Man, and
  Cybernetics: Systems 51~(2) (2019) 1097--1106.

\bibitem{xu2017distributed}
Y.~Xu, T.~Han, K.~Cai, Z.~Lin, G.~Yan, M.~Fu, A distributed algorithm for
  resource allocation over dynamic digraphs, IEEE Transactions on Signal
  Processing 65~(10) (2017) 2600--2612.

\bibitem{alghunaim2021dual}
S.~Alghunaim, Q.~Lyu, M.~Yan, A.~Sayed, Dual consensus proximal algorithm for
  multi-agent sharing problems, IEEE Transactions on Signal Processing 69
  (2021) 5568--5579.

\bibitem{li2022implicit}
J.~Li, H.~Su, Implicit tracking-based distributed constraint-coupled
  optimization, IEEE Transactions on Control of Network Systems (2022).

\bibitem{wu2022distributed}
W.~Wu, S.~Liu, S.~Zhu, Distributed dual gradient tracking for economic dispatch
  in power systems with noisy information, Electric Power Systems Research 211
  (2022) 108298.

\bibitem{zholbaryssov2022fast}
M.~Zholbaryssov, C.~Hadjicostis, A.~Dominguez-Garcia, Fast coordination of
  distributed energy resources over time-varying communication networks, IEEE
  Transactions on Automatic Control 68~(2) (2022) 1023--1038.

\bibitem{huang2024distributed}
Y.~Huang, Z.~Meng, J.~Sun, G.~Wang, Distributed continuous-time proximal
  algorithm for nonsmooth resource allocation problem with coupled constraints,
  Automatica 159 (2024) 111309.

\bibitem{10176349}
Y.~Huang, Z.~Meng, J.~Sun, W.~Ren, Distributed multi-proximal algorithm for
  nonsmooth convex optimization with coupled inequality constraints, IEEE
  Transactions on Automatic Control (2023) 1--8\href
  {https://doi.org/10.1109/TAC.2023.3293521}
  {\path{doi:10.1109/TAC.2023.3293521}}.

\bibitem{boyd2011distributed}
S.~Boyd, N.~Parikh, E.~Chu, B.~Peleato, J.~Eckstein, et~al., Distributed
  optimization and statistical learning via the alternating direction method of
  multipliers, Foundations and Trends in Machine learning 3~(1) (2011) 1--122.

\bibitem{iutzeler2015explicit}
F.~Iutzeler, P.~Bianchi, P.~Ciblat, W.~Hachem, Explicit convergence rate of a
  distributed alternating direction method of multipliers, IEEE Transactions on
  Automatic Control 61~(4) (2015) 892--904.

\bibitem{bertsekas2015parallel}
D.~Bertsekas, J.~Tsitsiklis, Parallel and distributed computation: numerical
  methods, Athena Scientific, 2015.

\bibitem{wei2012distributed}
E.~Wei, A.~Ozdaglar, Distributed alternating direction method of multipliers,
  in: IEEE 51st IEEE Conference on Decision and Control (CDC), IEEE, 2012, pp.
  5445--5450.

\bibitem{chang2016proximal}
T.~Chang, A proximal dual consensus {ADMM} method for multi-agent constrained
  optimization, IEEE Transactions on Signal Processing 64~(14) (2016)
  3719--3734.

\bibitem{wei_me_cdc}
W.~Jiang, M.~Doostmohammadian, T.~Charalambous, Distributed resource allocation
  via {ADMM} over digraphs, in: IEEE 61st Conference on Decision and Control
  (CDC), IEEE, 2022, pp. 5645--5651.

\bibitem{dtac}
M.~Doostmohammadian, W.~Jiang, T.~Charalambous, {DTAC-ADMM}: Delay-tolerant
  augmented consensus {ADMM}-based algorithm for distributed resource
  allocation, in: IEEE 61st Conference on Decision and Control (CDC), IEEE,
  2022, pp. 308--315.

\bibitem{falsone2018distributed}
A.~Falsone, K.~Margellos, M.~Prandini, A distributed iterative algorithm for
  multi-agent milps: Finite-time feasibility and performance characterization,
  IEEE control systems letters 2~(4) (2018) 563--568.

\bibitem{aybat2019distributed}
N.~Aybat, E.~Yazdandoost~Hamedani, A distributed {ADMM}-like method for
  resource sharing over time-varying networks, SIAM Journal on Optimization
  29~(4) (2019) 3036--3068.

\bibitem{makhdoumi2017convergence}
A.~Makhdoumi, A.~Ozdaglar, Convergence rate of distributed {ADMM} over
  networks, IEEE Transactions on Automatic Control 62~(10) (2017) 5082--5095.

\bibitem{chang2014multi}
T.~Chang, M.~Hong, X.~Wang, Multi-agent distributed optimization via inexact
  consensus {ADMM}, IEEE Transactions on Signal Processing 63~(2) (2014)
  482--497.

\bibitem{jian2019distributed}
L.~Jian, J.~Hu, J.~Wang, K.~Shi, Distributed inexact dual consensus {ADMM} for
  network resource allocation, Optimal Control Applications and Methods 40~(6)
  (2019) 1071--1087.

\bibitem{deng2017parallel}
W.~Deng, M.~Lai, Z.~Peng, W.~Yin, Parallel multi-block {ADMM} with {O(1/k)}
  convergence, Journal of Scientific Computing 71 (2017) 712--736.

\bibitem{yang2022proximal}
Y.~Yang, Q.~Jia, Z.~Xu, X.~Guan, C.~J. Spanos, Proximal {ADMM} for nonconvex
  and nonsmooth optimization, Automatica 146 (2022) 110551.

\bibitem{themis2021cpu}
A.~Grammenos, T.~Charalambous, E.~Kalyvianaki, {CPU} scheduling in data centers
  using asynchronous finite-time distributed coordination mechanisms, arXiv
  preprint arXiv:2101.06139 (2021).

\bibitem{chen2016distributed}
G.~Chen, J.~Ren, E.~N. Feng, Distributed finite-time economic dispatch of a
  network of energy resources, IEEE Transactions on Smart Grid 8~(2) (2016)
  822--832.

\bibitem{parsegov2013fixed}
S.~E. Parsegov, A.~E. Polyakov, P.~S. Shcherbakov, Fixed-time consensus
  algorithm for multi-agent systems with integrator dynamics, IFAC Proceedings
  Volumes 46~(27) (2013) 110--115.

\bibitem{garg2019fixed2}
K.~Garg, M.~Baranwal, A.~O. Hero, D.~Panagou, Fixed-time distributed
  optimization under time-varying communication topology, arXiv preprint
  arXiv:1905.10472 (2019).

\bibitem{wei2017consensus}
J.~Wei, A.~R.~F. Everts, M.~K. Camlibel, A.~J. van~der Schaft, Consensus
  dynamics with arbitrary sign-preserving nonlinearities, Automatica 83 (2017)
  226--233.

\bibitem{stankovic2019robust}
S.~S. Stankovi{\'c}, M.~Beko, M.~S. Stankovi{\'c}, Robust nonlinear consensus
  seeking, in: 58th IEEE Conference on Decision and Control, 2019, pp.
  4465--4470.

\bibitem{nedic2008distributed}
A.~Nedic, A.~Olshevsky, A.~Ozdaglar, J.~N. Tsitsiklis, Distributed subgradient
  methods and quantization effects, in: 47th IEEE Conference on Decision and
  Control, 2008, pp. 4177--4184.

\bibitem{rikos2020distributed}
A.~I. Rikos, C.~N. Hadjicostis, Distributed average consensus under quantized
  communication via event-triggered mass splitting, IFAC-PapersOnLine 53~(2)
  (2020) 2957--2962.

\bibitem{wei2018nonlinear}
J.~Wei, X.~Yi, H.~Sandberg, K.~H. Johansson, Nonlinear consensus protocols with
  applications to quantized communication and actuation, IEEE Transactions on
  Control of Network Systems 6~(2) (2019) 598--608.

\bibitem{nekouei2016convergence}
E.~Nekouei, T.~Alpcan, G.~N. Nair, R.~J. Evans, Convergence analysis of
  quantized primal-dual algorithms in network utility maximization problems,
  IEEE Transactions on Control of Network Systems 5~(1) (2016) 284--297.

\bibitem{nekouei2016performance}
E.~Nekouei, G.~N. Nair, T.~Alpcan, Performance analysis of gradient-based nash
  seeking algorithms under quantization, IEEE Transactions on Automatic Control
  61~(12) (2016) 3771--3783.

\bibitem{Hadjicostis_run:sum}
C.~N. Hadjicostis, N.~H. Vaidya, A.~D. Domínguez-García, Robust distributed
  average consensus via exchange of running sums, IEEE Transactions on
  Automatic Control 61~(6) (2016) 1492--1507.

\bibitem{liu2020global}
Z.~Liu, A.~Saberi, A.~A. Stoorvogel, D.~Nojavanzadeh, Global regulated state
  synchronization for homogeneous networks of non-introspective agents in
  presence of input saturation: Scale-free nonlinear and linear protocol
  designs, Automatica 119 (2020) 109041.

\bibitem{vtc}
M.~Doostmohammadian, M.~Vrakopoulou, A.~Aghasi, T.~Charalambous, Distributed
  finite-sum constrained optimization subject to nonlinearity on the node
  dynamics, in: IEEE 95th Vehicular Technology Conference:(VTC2022-Spring),
  IEEE, 2022, pp. 1--6.

\bibitem{dsvm2}
M.~Doostmohammadian, A.~Aghasi, H.~Zarrabi, {D-SVM} over networked systems with
  non-ideal linking conditions, Iran Journal of Computer Science (2023) 1--12.

\bibitem{xin2018linear}
R.~Xin, U.~A. Khan, A linear algorithm for optimization over directed graphs
  with geometric convergence, IEEE Control Systems Letters 2~(3) (2018)
  315--320.

\bibitem{ling2013decentralized}
Q.~Ling, A.~Ribeiro, Decentralized dynamic optimization through the alternating
  direction method of multipliers, IEEE Transactions on Signal Processing
  62~(5) (2013) 1185--1197.

\bibitem{nedic2017achieving}
A.~Nedic, A.~Olshevsky, W.~Shi, Achieving geometric convergence for distributed
  optimization over time-varying graphs, SIAM Journal on Optimization 27~(4)
  (2017) 2597--2633.

\bibitem{scaman2017optimal}
K.~Scaman, F.~Bach, S.~Bubeck, Y.~Lee, L.~Massouli{\'e}, Optimal algorithms for
  smooth and strongly convex distributed optimization in networks, in:
  International conference on machine learning, PMLR, 2017, pp. 3027--3036.

\bibitem{olshevsky2015linear}
A.~Olshevsky, Linear time average consensus on fixed graphs, IFAC-PapersOnLine
  48~(22) (2015) 94--99.

\bibitem{nedic2017geometrically}
A.~Nedi{\'c}, A.~Olshevsky, W.~Shi, C.~Uribe, Geometrically convergent
  distributed optimization with uncoordinated step-sizes, in: American Control
  Conference (ACC), IEEE, 2017, pp. 3950--3955.

\bibitem{stich2019local}
S.~Stich, Local {SGD} converges fast and communicates little, in:
  2019-International Conference on Learning Representations, 2019.

\bibitem{haddadpour2019trading}
F.~Haddadpour, M.~Kamani, M.~Mahdavi, V.~Cadambe, Trading redundancy for
  communication: Speeding up distributed sgd for non-convex optimization, in:
  International Conference on Machine Learning, PMLR, 2019, pp. 2545--2554.

\bibitem{Kar6345156}
S.~Kar, G.~Hug, Distributed robust economic dispatch in power systems: A
  consensus + innovations approach, in: IEEE Power and Energy Society General
  Meeting, 2012, pp. 1--8.

\bibitem{aybat2016primal}
N.~Aybat, E.~Yazdandoost~Hamedani, A primal-dual method for conic constrained
  distributed optimization problems, Advances in neural information processing
  systems 29 (2016).

\bibitem{beck20141}
A.~Beck, A.~Nedi{\'c}, A.~Ozdaglar, M.~Teboulle, An $ o (1/k) $ gradient method
  for network resource allocation problems, IEEE Transactions on Control of
  Network Systems 1~(1) (2014) 64--73.

\bibitem{zargham2013accelerated}
M.~Zargham, A.~Ribeiro, A.~Ozdaglar, A.~Jadbabaie, Accelerated dual descent for
  network flow optimization, IEEE Transactions on Automatic Control 59~(4)
  (2013) 905--920.

\bibitem{xin2019distributed}
R.~Xin, A.~K. Sahu, U.~A. Khan, S.~Kar, Distributed stochastic optimization
  with gradient tracking over strongly-connected networks, in: IEEE Conference
  on Decision and Control (CDC), 2019, pp. 8353--8358.

\bibitem{dauphin2015equilibrated}
Y.~Dauphin, H.~De~Vries, Y.~Bengio, Equilibrated adaptive learning rates for
  non-convex optimization, Advances in neural information processing systems 28
  (2015).

\bibitem{sayedchen11}
J.~Chen, A.~H. Sayed, Diffusion adaptation strategies for distributed
  optimization and learning over networks, IEEE Transactions on Signal
  Processing, 60~(8) (2012) 4289--4305.

\bibitem{chen2011distributed}
J.~Chen, S.-Y. Tu, A.~H. Sayed, Distributed optimization via diffusion
  adaptation, in: 4th IEEE International Workshop on Computational Advances in
  Multi-Sensor Adaptive Processing (CAMSAP), IEEE, 2011, pp. 281--284.

\bibitem{liuvariance}
L.~Liu, H.~Jiang, P.~He, W.~Chen, X.~Liu, J.~Gao, J.~Han, On the variance of
  the adaptive learning rate and beyond, in: International Conference on
  Learning Representations, 2019.

\bibitem{li2020communication}
B.~Li, S.~Cen, Y.~Chen, Y.~Chi, Communication-efficient distributed
  optimization in networks with gradient tracking and variance reduction, The
  Journal of Machine Learning Research 21~(1) (2020) 7331--7381.

\bibitem{ghadikolaei2020communication}
H.~S. Ghadikolaei, S.~Magn{\'u}sson, Communication-efficient variance-reduced
  stochastic gradient descent, IFAC-PapersOnLine 53~(2) (2020) 2648--2653.

\bibitem{qureshi2021push}
M.~Qureshi, R.~Xin, S.~Kar, U.~A. Khan, {Push-SAGA}: A decentralized stochastic
  algorithm with variance reduction over directed graphs, IEEE Control Systems
  Letters 6 (2021) 1202--1207.

\bibitem{wang2020distributed}
B.~Wang, Q.~Fei, Q.~Wu, Distributed time-varying resource allocation
  optimization based on finite-time consensus approach, IEEE Control Systems
  Letters 5~(2) (2020) 599--604.

\bibitem{li2020distributed2}
W.~Li, Z.~Lin, K.~Cai, G.~Yan, Distributed algorithm for a finite time horizon
  resource allocation over a directed network, IET Control Theory \&
  Applications 14~(9) (2020) 1170--1182.

\bibitem{dai2022consensus}
H.~Dai, X.~Fang, J.~Jia, Consensus-based distributed fixed-time optimization
  for a class of resource allocation problems, Journal of the Franklin
  Institute 359~(18) (2022) 11135--11154.

\bibitem{song2021fixed}
Y.~Song, J.~Cao, L.~Rutkowski, A fixed-time distributed optimization algorithm
  based on event-triggered strategy, IEEE Transactions on Network Science and
  Engineering 9~(3) (2021) 1154--1162.

\bibitem{chen2020initialization}
G.~Chen, Z.~Guo, Initialization-free distributed fixed-time convergent
  algorithms for optimal resource allocation, IEEE Transactions on Systems,
  Man, and Cybernetics: Systems 52~(2) (2020) 845--854.

\bibitem{shi2022distributed}
X.~Shi, L.~Xu, T.~Yang, Z.~Lin, X.~Wang, Distributed fixed-time resource
  allocation algorithm for the general linear multi-agent systems, IEEE
  Transactions on Circuits and Systems II: Express Briefs 69~(6) (2022)
  2867--2871.

\bibitem{li2020distributed3}
Z.~Li, Z.~Ding, Distributed multiobjective optimization for network resource
  allocation of multiagent systems, IEEE Transactions on Cybernetics 51~(12)
  (2020) 5800--5810.

\bibitem{guo2022distributed}
Z.~Guo, G.~Chen, Distributed dynamic event-triggered and practical
  predefined-time resource allocation in cyber--physical systems, Automatica
  142 (2022) 110390.

\bibitem{9609591}
X.~Gong, Y.~Cui, J.~Shen, J.~Xiong, T.~Huang, Distributed optimization in
  prescribed-time: Theory and experiment, IEEE Transactions on Network Science
  and Engineering 9~(2) (2022) 564--576.

\bibitem{lin2020predefined}
W.~Lin, Y.~Wang, C.~Li, X.~Yu, Predefined-time optimization for distributed
  resource allocation, Journal of the Franklin Institute 357~(16) (2020)
  11323--11348.

\bibitem{wang2019distributed}
X.~Wang, Y.~Hong, X.~Sun, K.~Liu, Distributed optimization for resource
  allocation problems under large delays, IEEE Transactions on Industrial
  Electronics 66~(12) (2019) 9448--9457.

\bibitem{kia2017distributed}
S.~S. Kia, Distributed optimal in-network resource allocation algorithm design
  via a control theoretic approach, Systems \& Control Letters 107 (2017)
  49--57.

\bibitem{lin2021distributed}
W.~Lin, Y.~Wang, C.~Li, X.~Yu, Distributed resource allocation via accelerated
  saddle point dynamics, IEEE/CAA Journal of Automatica Sinica 8~(9) (2021)
  1588--1599.

\bibitem{icrom}
M.~Doostmohammadian, U.~A. Khan, A.~Aghasi, Distributed constraint-coupled
  optimization over unreliable networks, in: 10th RSI International Conference
  on Robotics and Mechatronics (ICRoM), IEEE, 2022, pp. 371--376.

\bibitem{li2016consensus}
L.~Li, D.~Ho, J.~Lu, A consensus recovery approach to nonlinear multi-agent
  system under node failure, Information Sciences 367 (2016) 975--989.

\bibitem{sundaram2016secure}
S.~Sundaram, B.~Gharesifard, Secure local filtering algorithms for distributed
  optimization, in: IEEE 55th Conference on Decision and Control (CDC), IEEE,
  2016, pp. 1871--1876.

\bibitem{falsone2022sensitivity}
A.~Falsone, K.~Margellos, J.~Zizzo, M.~Prandini, S.~Garatti, On the sensitivity
  of linear resource sharing problems to the arrival of new agents, IEEE
  Transactions on Automatic Control (2022).

\bibitem{fu2021resilient}
W.~Fu, Q.~Ma, J.~Qin, Y.~Kang, Resilient consensus-based distributed
  optimization under deception attacks, International Journal of Robust and
  Nonlinear Control 31~(6) (2021) 1803--1816.

\bibitem{shao2020distributed}
G.~Shao, R.~Wang, X.~Wang, K.~Liu, Distributed algorithm for resource
  allocation problems under persistent attacks, Journal of the Franklin
  Institute 357~(10) (2020) 6241--6256.

\bibitem{sardana2009auto}
A.~Sardana, R.~Joshi, An auto-responsive honeypot architecture for dynamic
  resource allocation and qos adaptation in ddos attacked networks, Computer
  Communications 32~(12) (2009) 1384--1399.

\bibitem{turan2020resilient}
B.~Turan, C.~A. Uribe, H.~Wai, M.~Alizadeh, Resilient primal--dual optimization
  algorithms for distributed resource allocation, IEEE Transactions on Control
  of Network Systems 8~(1) (2020) 282--294.

\bibitem{xu2022resilient}
C.~Xu, Q.~Liu, T.~Huang, Resilient penalty function method for distributed
  constrained optimization under byzantine attack, Information Sciences 596
  (2022) 362--379.

\bibitem{wang2022byzantine}
R.~Wang, Y.~Liu, Q.~Ling, Byzantine-resilient resource allocation over
  decentralized networks, IEEE Transactions on Signal Processing 70 (2022)
  4711--4726.

\bibitem{XU2022}
C.~Xu, Q.~Liu, A resilient distributed optimization algorithm based on
  consensus of multi-agent system against two attack scenarios, Journal of the
  Franklin Institute (2022).

\bibitem{yemini2022resilience}
M.~Yemini, A.~Nedi{\'c}, S.~Gil, A.~J. Goldsmith, Resilience to malicious
  activity in distributed optimization for cyberphysical systems, in: IEEE 61st
  Conference on Decision and Control (CDC), IEEE, 2022, pp. 4185--4192.

\bibitem{beaude2020privacy}
O.~Beaude, P.~Benchimol, S.~Gaubert, P.~Jacquot, N.~Oudjane, A
  privacy-preserving method to optimize distributed resource allocation, SIAM
  Journal on Optimization 30~(3) (2020) 2303--2336.

\bibitem{angel2020private}
S.~Angel, S.~Kannan, Z.~Ratliff, Private resource allocators and their
  applications, in: IEEE Symposium on Security and Privacy, IEEE, 2020, pp.
  372--391.

\bibitem{lin2020distributed}
W.~Lin, Y.~Wang, C.~Li, X.~Yu, Distributed resource allocation: an indirect
  dual ascent method with an exponential convergence rate, Nonlinear Dynamics
  102 (2020) 1685--1699.

\bibitem{chen2021distributed}
G.~Chen, Z.~Li, Distributed optimal resource allocation over strongly connected
  digraphs: A surplus-based approach, Automatica 125 (2021) 109459.

\bibitem{jacquot2019privacy}
P.~Jacquot, O.~Beaude, P.~Benchimol, S.~Gaubert, N.~Oudjane, A
  privacy-preserving disaggregation algorithm for non-intrusive management of
  flexible energy, in: IEEE 58th Conference on Decision and Control (CDC),
  IEEE, 2019, pp. 890--896.

\bibitem{nedic2009distributed}
A.~Nedic, A.~Ozdaglar, Distributed subgradient methods for multi-agent
  optimization, IEEE Transactions on Automatic Control 54~(1) (2009) 48--61.

\bibitem{derm}
M.~Doostmohammadian, Distributed energy resource management: All-time
  resource-demand feasibility, delay-tolerance, nonlinearity, and beyond, IEEE
  Control Systems Letters (2023).

\bibitem{scl}
M.~Doostmohammadian, A.~Aghasi, M.~Vrakopoulou, H.~R. Rabiee, U.~A. Khan,
  T.~Charalambous, Distributed delay-tolerant strategies for
  equality-constraint sum-preserving resource allocation, Systems \& Control
  Letters (2023).

\bibitem{seuret2008consensus}
A.~Seuret, D.~Dimarogonas, K.~H. Johansson, Consensus under communication
  delays, in: 47th IEEE Conference on Decision and Control, IEEE, 2008, pp.
  4922--4927.

\bibitem{Themis_delay}
C.~N. {Hadjicostis}, T.~{Charalambous}, Average consensus in the presence of
  delays in directed graph topologies, IEEE Transactions on Automatic Control
  59~(3) (2014) 763--768.

\bibitem{acc22delay}
M.~Doostmohammadian, U.~Khan, M.~Pirani, T.~Charalambous, Consensus-based
  distributed estimation in the presence of heterogeneous, time-invariant
  delays, IEEE Control Systems Letters 6 (2021) 1598--1603.

\bibitem{schenato2011average}
L.~Schenato, F.~Fiorentin, Average timesynch: A consensus-based protocol for
  clock synchronization in wireless sensor networks, Automatica 47~(9) (2011)
  1878--1886.

\bibitem{schenato2007distributed}
L.~Schenato, G.~Gamba, A distributed consensus protocol for clock
  synchronization in wireless sensor network, in: 46th ieee conference on
  decision and control, IEEE, 2007, pp. 2289--2294.

\bibitem{chen2017delay}
G.~Chen, Z.~Zhao, Delay effects on consensus-based distributed economic
  dispatch algorithm in microgrid, IEEE Transactions on Power Systems 33~(1)
  (2017) 602--612.

\bibitem{fagnani2009average}
F.~Fagnani, S.~Zampieri, Average consensus with packet drop communication, SIAM
  Journal on Control and Optimization 48~(1) (2009) 102--133.

\bibitem{vaidya2012robust}
N.~Vaidya, C.~Hadjicostis, A.~Dom{\'\i}nguez-Garc{\'\i}a, Robust average
  consensus over packet dropping links: Analysis via coefficients of
  ergodicity, in: IEEE 51st IEEE Conference on Decision and Control (CDC),
  IEEE, 2012, pp. 2761--2766.

\bibitem{liu2015leader}
Z.~Liu, X.~You, H.~Yang, L.~Zhao, Leader-following consensus of heterogeneous
  multi-agent systems with packet dropout, International Journal of Control,
  Automation and Systems 13 (2015) 1067--1075.

\bibitem{hadjicostis2015robust}
C.~Hadjicostis, N.~Vaidya, A.~Dom{\'\i}nguez-Garc{\'\i}a, Robust distributed
  average consensus via exchange of running sums, IEEE Transactions on
  Automatic Control 61~(6) (2015) 1492--1507.

\bibitem{xia2010optimal}
X.~Xia, A.~M. Elaiw, Optimal dynamic economic dispatch of generation: A review,
  Electric power systems research 80~(8) (2010) 975--986.

\bibitem{chowdhury1990review}
B.~Chowdhury, S.~Rahman, A review of recent advances in economic dispatch, IEEE
  Transactions on Power Systems 5~(4) (1990) 1248--1259.

\bibitem{elsayed2014fully}
W.~Elsayed, E.~El-Saadany, A fully decentralized approach for solving the
  economic dispatch problem, IEEE Transactions on power systems 30~(4) (2014)
  2179--2189.

\bibitem{firouzbahrami2022finite}
M.~Firouzbahrami, A.~Nobakhti, Finite-time distributed economic dispatch over
  network systems with coupled local costs, IEEE Control Systems Letters 7
  (2022) 325--330.

\bibitem{cherukuri2014distributed}
A.~Cherukuri, S.~Mart{\'\i}nez, J.~Cort{\'e}s, Distributed, anytime
  optimization in power-generator networks for economic dispatch, in: American
  Control Conference, IEEE, 2014, pp. 172--177.

\bibitem{pourbabak2017novel}
H.~Pourbabak, J.~Luo, T.~Chen, W.~Su, A novel consensus-based distributed
  algorithm for economic dispatch based on local estimation of power mismatch,
  IEEE Transactions on Smart Grid 9~(6) (2017) 5930--5942.

\bibitem{rikos2022distributed}
A.~I. Rikos, J.~Nyl{\"o}f, S.~Gracy, K.~H. Johansson, Distributed optimal
  allocation with quantized communication and privacy-preserving guarantees,
  IFAC-PapersOnLine 55~(41) (2022) 64--70.

\bibitem{grammenos2023cpu}
A.~Grammenos, T.~Charalambous, E.~Kalyvianaki, {CPU} scheduling in data centers
  using asynchronous finite-time distributed coordination mechanisms, IEEE
  Transactions on Network Science and Engineering (2023).

\bibitem{kalyvianaki2009self}
E.~Kalyvianaki, T.~Charalambous, S.~Hand, Self-adaptive and self-configured
  {CPU} resource provisioning for virtualized servers using kalman filters, in:
  Proceedings of the 6th International Conference on Autonomic Computing, 2009,
  pp. 117--126.

\bibitem{makridis2020robust}
E.~Makridis, K.~Deliparaschos, E.~Kalyvianaki, A.~Zolotas, T.~Charalambous,
  Robust dynamic {CPU} resource provisioning in virtualized servers, IEEE
  Transactions on Services Computing 15~(2) (2020) 956--969.

\bibitem{chen2022resource}
X.~Chen, L.~Yang, Z.~Chen, G.~Min, X.~Zheng, C.~Rong, Resource allocation with
  workload-time windows for cloud-based software services: a deep reinforcement
  learning approach, IEEE Transactions on Cloud Computing (2022).

\bibitem{hattab2019optimized}
G.~Hattab, S.~Ucar, T.~Higuchi, O.~Altintas, F.~Dressler, D.~Cabric, Optimized
  assignment of computational tasks in vehicular micro clouds, in: 2nd
  International Workshop on Edge Systems, Analytics and Networking, 2019, pp.
  1--6.

\bibitem{mann2015allocation}
Z.~Mann, Allocation of virtual machines in cloud data centers--a survey of
  problem models and optimization algorithms, {ACM} Computing Surveys (CSUR)
  48~(1) (2015) 1--34.

\bibitem{li2015connecting}
N.~Li, C.~Zhao, L.~Chen, Connecting automatic generation control and economic
  dispatch from an optimization view, IEEE Transactions on Control of Network
  Systems 3~(3) (2015) 254--264.

\bibitem{kumar2005recent}
P.~Kumar, D.~Kothari, Recent philosophies of automatic generation control
  strategies in power systems, IEEE transactions on power systems 20~(1) (2005)
  346--357.

\bibitem{ullah2021automatic}
K.~Ullah, A.~Basit, Z.~Ullah, S.~Aslam, H.~Herodotou, Automatic generation
  control strategies in conventional and modern power systems: A comprehensive
  overview, Energies 14~(9) (2021) 2376.

\bibitem{bevrani2017intelligent}
H.~Bevrani, T.~Hiyama, Intelligent automatic generation control, CRC press,
  2017.

\bibitem{zhou2020distributed}
Q.~Zhou, M.~Shahidehpour, A.~Paaso, S.~Bahramirad, A.~Alabdulwahab,
  A.~Abusorrah, Distributed control and communication strategies in networked
  microgrids, IEEE Communications Surveys \& Tutorials 22~(4) (2020)
  2586--2633.

\bibitem{wood2013power}
A.~J. Wood, B.~F. Wollenberg, G.~B. Shebl{\'e}, Power generation, operation,
  and control, John Wiley \& Sons, 2013.

\bibitem{sun2017optimal}
S.~Sun, Q.~Yang, W.~Yan, Optimal temporal-spatial {PEV} charging scheduling in
  active power distribution networks, Protection and Control of Modern Power
  Systems 2 (2017) 1--10.

\bibitem{kisacikoglu2017distributed}
M.~Kisacikoglu, F.~Erden, N.~Erdogan, Distributed control of {PEV} charging
  based on energy demand forecast, IEEE Transactions on Industrial Informatics
  14~(1) (2017) 332--341.

\bibitem{mukherjee2016distributed}
J.~Mukherjee, A.~Gupta, Distributed charge scheduling of plug-in electric
  vehicles using inter-aggregator collaboration, IEEE Transactions on Smart
  Grid 8~(1) (2016) 331--341.

\bibitem{li2019decentralized}
M.~Li, J.~Gao, N.~Chen, L.~Zhao, X.~Shen, Decentralized {PEV} power allocation
  with power distribution and transportation constraints, IEEE Journal on
  Selected Areas in Communications 38~(1) (2019) 229--243.

\bibitem{abdalrahman2017survey}
A.~Abdalrahman, W.~Zhuang, A survey on {PEV} charging infrastructure: Impact
  assessment and planning, Energies 10~(10) (2017) 1650.

\bibitem{dallinger2012grid}
D.~Dallinger, M.~Wietschel, Grid integration of intermittent renewable energy
  sources using price-responsive plug-in electric vehicles, Renewable and
  Sustainable Energy Reviews 16~(5) (2012) 3370--3382.

\bibitem{amin2020review}
A.~Amin, W.~Tareen, M.~Usman, H.~Ali, I.~Bari, B.~Horan, S.~Mekhilef, M.~Asif,
  S.~Ahmed, A.~Mahmood, A review of optimal charging strategy for electric
  vehicles under dynamic pricing schemes in the distribution charging network,
  Sustainability 12~(23) (2020) 10160.

\bibitem{esmaili2014optimal}
M.~Esmaili, M.~Rajabi, Optimal charging of plug-in electric vehicles observing
  power grid constraints, IET Generation, Transmission \& Distribution 8~(4)
  (2014) 583--590.

\bibitem{falsone2017dual}
A.~Falsone, K.~Margellos, S.~Garatti, M.~Prandini, Dual decomposition for
  multi-agent distributed optimization with coupling constraints, Automatica 84
  (2017) 149--158.

\bibitem{vujanic2016decomposition}
R.~Vujanic, P.~M. Esfahani, P.~J. Goulart, S.~Mari{\'e}thoz, M.~Morari, A
  decomposition method for large scale milps, with performance guarantees and a
  power system application, Automatica 67 (2016) 144--156.

\bibitem{deori2016decentralized}
L.~Deori, K.~Margellos, M.~Prandini, On decentralized convex optimization in a
  multi-agent setting with separable constraints and its application to optimal
  charging of electric vehicles, in: IEEE 55th Conference on Decision and
  Control (CDC), IEEE, 2016, pp. 6044--6049.

\bibitem{ioan2021mixed}
D.~Ioan, I.~Prodan, S.~Olaru, F.~Stoican, S.~Niculescu, Mixed-integer
  programming in motion planning, Annual Reviews in Control 51 (2021) 65--87.

\bibitem{li2020distributed}
X.~Li, X.~Yi, L.~Xie, Distributed online optimization for multi-agent networks
  with coupled inequality constraints, IEEE Transactions on Automatic Control
  66~(8) (2020) 3575--3591.

\bibitem{chiang2007layering}
M.~Chiang, S.~H. Low, A.~R. Calderbank, J.~C. Doyle, Layering as optimization
  decomposition: A mathematical theory of network architectures, Proceedings of
  the IEEE 95~(1) (2007) 255--312.

\bibitem{palomar2006tutorial}
D.~P. Palomar, M.~Chiang, A tutorial on decomposition methods for network
  utility maximization, IEEE Journal on Selected Areas in Communications 24~(8)
  (2006) 1439--1451.

\bibitem{lin2006tutorial}
X.~Lin, N.~B. Shroff, R.~Srikant, A tutorial on cross-layer optimization in
  wireless networks, IEEE Journal on Selected areas in Communications 24~(8)
  (2006) 1452--1463.

\bibitem{srikant2013communication}
R.~Srikant, L.~Ying, Communication networks: an optimization, control, and
  stochastic networks perspective, Cambridge University Press, 2013.

\bibitem{hasan2016network}
N.~Hasan, W.~Ejaz, N.~Ejaz, H.~S. Kim, A.~Anpalagan, M.~Jo, Network selection
  and channel allocation for spectrum sharing in 5g heterogeneous networks,
  IEEE Access 4 (2016) 980--992.

\bibitem{tsiropoulos2014radio}
G.~I. Tsiropoulos, O.~A. Dobre, M.~H. Ahmed, K.~E. Baddour, Radio resource
  allocation techniques for efficient spectrum access in cognitive radio
  networks, IEEE Communications Surveys \& Tutorials 18~(1) (2014) 824--847.

\bibitem{kraning2013dynamic}
M.~Kraning, E.~Chu, J.~Lavaei, S.~Boyd, Dynamic network energy management via
  proximal message passing, Foundations and Trends{\textregistered} in
  Optimization 1~(2) (2013) 73--126.

\bibitem{trossen2012designing}
D.~Trossen, G.~Parisis, Designing and realizing an information-centric
  internet, IEEE Communications Magazine 50~(7) (2012) 60--67.

\bibitem{ercetin2003market}
O.~Ercetin, L.~Tassiulas, Market-based resource allocation for content delivery
  in the internet, IEEE Transactions on Computers 52~(12) (2003) 1573--1585.

\bibitem{guo2018energy}
S.~Guo, J.~Liu, Y.~Yang, B.~Xiao, Z.~Li, Energy-efficient dynamic computation
  offloading and cooperative task scheduling in mobile cloud computing, IEEE
  Transactions on Mobile Computing 18~(2) (2018) 319--333.

\bibitem{ergu2013analytic}
D.~Ergu, G.~Kou, Y.~Peng, Y.~Shi, Y.~Shi, The analytic hierarchy process: task
  scheduling and resource allocation in cloud computing environment, The
  Journal of Supercomputing 64 (2013) 835--848.

\bibitem{chiang2016fog}
M.~Chiang, T.~Zhang, Fog and iot: An overview of research opportunities, IEEE
  Internet of things journal 3~(6) (2016) 854--864.

\bibitem{kelly1998rate}
F.~P. Kelly, A.~K. Maulloo, D.~K.~H. Tan, Rate control for communication
  networks: shadow prices, proportional fairness and stability, Journal of the
  Operational Research society 49 (1998) 237--252.

\bibitem{enyioha2015distributed}
C.~Enyioha, A.~Jadbabaie, V.~Preciado, G.~Pappas, Distributed resource
  allocation for control of spreading processes, in: 2015 European Control
  Conference (ECC), IEEE, 2015, pp. 2216--2221.

\bibitem{nowzari2016analysis}
C.~Nowzari, V.~M. Preciado, G.~J. Pappas, Analysis and control of epidemics: A
  survey of spreading processes on complex networks, IEEE Control Systems
  Magazine 36~(1) (2016) 26--46.

\bibitem{peteiro2013survey}
D.~Peteiro-Barral, B.~Guijarro-Berdi{\~n}as, A survey of methods for
  distributed machine learning, Progress in Artificial Intelligence 2 (2013)
  1--11.

\bibitem{kumar2022machine}
Y.~Kumar, S.~Kaul, Y.~Hu, Machine learning for energy-resource allocation,
  workflow scheduling and live migration in cloud computing: State-of-the-art
  survey, Sustainable Computing: Informatics and Systems 36 (2022) 100780.

\bibitem{mohajer2017big}
A.~Mohajer, M.~Barari, H.~Zarrabi, Big data based self-optimization networking:
  A novel approach beyond cognition, Intelligent Automation \& Soft Computing
  (2017) 1--7.

\end{thebibliography}

\end{document}